\documentclass[a4paper,fleqn,usenatbib]{mnras}

\usepackage{newtxtext,newtxmath}
\usepackage[T1]{fontenc}
\usepackage{ae,aecompl}

\usepackage{graphicx}	% Including figure files
\usepackage{gensymb}
\usepackage{amsmath}	% Advanced maths commands
\usepackage{amssymb}	% Extra maths symbols
\usepackage{multirow}
\usepackage{pbox}
\usepackage{placeins}
\usepackage{easy-todo}	% Add comments and notes
\usepackage{pifont}

\usepackage{color}

\usepackage{soul} 
\usepackage{amsmath}
\usepackage{ulem}
\usepackage{xspace}

\newcommand{\WMAPonep}{\ensuremath{{\Ho{}=\kms[73]\pMpc{}},\,\allowbreak \Omega_M=0.25,\, \Omega_\Lambda=0.75,\, n_s=1.0,\, \sigma_8=0.9}}

\newcommand{\LowIncGammaFunc}[2]{\ensuremath{\gamma\left(#1,#2\right)}}

\newcommand{\appref}[1]{Appendix~\ref{#1}}
\newcommand{\secref}[1]{Section~\ref{#1}}
\newcommand{\tabref}[1]{Table~\ref{#1}}
\newcommand{\tabrefs}[2]{Tables~\ref{#1} and \ref{#2}}
\newcommand{\figref}[1]{Fig.~\ref{#1}}
\newcommand{\figrefs}[2]{Figs~\ref{#1}~and~\ref{#2}}
\newcommand{\eqnref}[1]{equation~(\ref{#1})}

\defcitealias{tollerud_hundreds_2008}{T08}
\defcitealias{koposov_luminosity_2008}{K08}
\defcitealias{walsh_invisibles_2009}{W09}
\defcitealias{jethwa_magellanic_2016}{J16}

\newcommand{\fitalpha}{\ensuremath{0.24}}
\newcommand{\fitconc}{\ensuremath{4.9}}

\newcommand{\origmedSDSSDESresult}{\ensuremath{141^{+54}_{-35}}}

\newcommand{\TtwomedSDSSDESresult}{\ensuremath{113^{+34}_{-24}}}

\newcommand{\medhfresult}{\ensuremath{61^{+37}_{-23}}}       % For 1.0e12 halo, 1sigma
\newcommand{\medufresult}{\ensuremath{46^{+12}_{-8}}}         % For 1.0e12 halo, 1sigma
\newcommand{\medSDSSDESresult}{\ensuremath{124^{+40}_{-27}}}   % For 1.0e12 halo, 1sigma

\newcommand{\tollerudresult}{\ensuremath{322^{+144}_{-76}}}       % Tollerud result within 300 kpc, 98%
\newcommand{\hargisufresult}{\ensuremath{66^{+9}_{-7}}}         % Hargis result within 300 kpc, 1sigma
\newcommand{\medSDSSDEStollimresult}{\ensuremath{66^{+39}_{-20}}} % Our result within 300 kpc, for 1.0e12 halo, for MV <= -2.7, 98% CL.
\newcommand{\medSDSStollimresult}{\ensuremath{64^{+55}_{-26}}} % Our SDSS-only result within 300 kpc, for 1.0e12 halo, for MV <= -2.7, 98% CL - note that this is an average between the two adjacent bins.
\newcommand{\tollresultfrac}{\ensuremath{5}} % The ratio of \tollerudresult and \medSDSStollimresult.

\newcommand{\Eagle}{\mbox{\normalfont{\sc eagle}}}
\newcommand{\Galform}{\mbox{\normalfont{\sc galform}}}
\newcommand{\GadgetIII}{\mbox{\normalfont{\sc p-gadget3}}}
\newcommand{\numpy}{{\sc numpy}}
\newcommand{\scipy}{{\sc scipy}}
\newcommand{\matplotlib}{{\sc matplotlib}}

\newcommand{\Apostle}{\mbox{\normalfont{\sc APOSTLE}}}
\newcommand{\Aquarius}{\mbox{\normalfont{\sc aquarius}}}

\newcommand{\conc}{\ensuremath{c_{200}}}
\newcommand{\LCDM}{\normalfont{$\Lambda$CDM}}
\newcommand{\LMC}{\normalfont{LMC}}
\newcommand{\mV}{\ensuremath{\rm m_V}}
\newcommand{\MV}{\ensuremath{M_{\rm V}}}
\newcommand{\Mvir}{\ensuremath{M_{\rm 200}}}
\newcommand{\Rout}{\kpc[300]}
\newcommand{\Rvir}{\ensuremath{\rm R_{200}}}
\newcommand{\CL}[1]{\ensuremath{#1}~CL}

\newcommand{\ATLAS}{\normalfont{VLT~ATLAS}}
\newcommand{\DES}{\normalfont{DES}}
\newcommand{\DR}[1]{\normalfont{DR{#1}}}
\newcommand{\HSC}{\normalfont{HSC}}
\newcommand{\LSST}{\normalfont{LSST}}
\newcommand{\Mag}{\normalfont{MagLiteS}}
\newcommand{\PAN}{\normalfont{Pan-STARRS}}

\newcommand{\SDSS}{\normalfont{SDSS}}
\newcommand{\SDSSDR}[1]{\normalfont{SDSS~DR{#1}}}
\newcommand{\SMASH}{\normalfont{SMASH}}
\newcommand{\twomass}{\normalfont{2MASS}}
\newcommand{\WMAP}{\normalfont{WMAP}}

\newcommand{\unit}[1]{\ensuremath{\mathrm{\,#1}}\xspace}
\newcommand{\kms}[1][]{\def\tst{#1}\ifx\tst\empty{\unit{km\, s^{-1}}}\else{\ensuremath{{#1}\, \unit{km\, s^{-1}}}}\fi}
\newcommand{\kpc}[1][]{\def\tst{#1}\ifx\tst\empty{\unit{kpc}}\else{\ensuremath{{#1}\, \unit{kpc}}}\fi}
\newcommand{\mgn}[1]{\ensuremath{{#1}\, \unit{mag}}}
\newcommand{\Mpc}{\unit{Mpc}}
\newcommand{\Gyr}[1]{\def\tst{#1}\ifx\tst\empty{\ensuremath{\unit{Gyr}}}\else{\ensuremath{{#1}\, \unit{Gyr}}}\fi}
\newcommand{\Msun}[1][]{\def\tst{#1}\ifx\tst\empty{\unit{M_\odot}}\else{\ensuremath{{#1}\, \unit{M_\odot}}}\fi}
\newcommand{\pc}{\unit{pc}}
\newcommand{\pMpc}{\unit{Mpc^{-1}}}

\renewcommand{\d}[1]{\ensuremath{\operatorname{d}\!{#1}}}
\newcommand{\Ho}{\ensuremath{H_0}}
\newcommand{\Reff}{\ensuremath{R_{\rm eff}}}
\newcommand{\router}{\ensuremath{R_\mathrm{out}}}
\newcommand{\vpeak}{\ensuremath{v_\mathrm{peak}}}
\title[The MW satellite galaxy population]{The total satellite population of the Milky Way}

\author[O. Newton et al.]{Oliver Newton\thanks{E-mail: oliver.j.newton@durham.ac.uk},
Marius Cautun,
Adrian Jenkins,
Carlos S. Frenk and
\newauthor
John C. Helly
\\
Institute for Computational Cosmology, Durham University, South Road, Durham, DH1 3LE, UK\\
}

\date{Accepted 2018 April 20. Received 2018 March 22; in original form 2017 August 14}

\pubyear{2018}

\begin{document}
\label{firstpage}
\pagerange{\pageref{firstpage}--\pageref{lastpage}}
\maketitle

\begin{abstract}
  The total number and luminosity function of the population of dwarf
  galaxies of the Milky Way (MW) provide important constraints on the
  nature of the dark matter and on the astrophysics of galaxy
  formation at low masses. However, only a partial census of this
  population exists because of the flux limits and restricted sky
  coverage of existing Galactic surveys. We combine the sample of satellites
  recently discovered by the Dark Energy Survey (\DES{}) with the
  satellites found in Sloan Digital Sky Survey (\SDSS{}) Data Release 9
  (together these surveys cover nearly half the sky) to estimate the total
  luminosity function of satellites down to $\MV{}=0$. We apply a new
  Bayesian inference method in which we assume that the radial
  distribution   of satellites independently of absolute magnitude
  follows that of   subhaloes selected according to their
  peak maximum circular velocity.  We find that there should be at least
  \medSDSSDESresult{}~(\CL{68\ {\rm per\ cent}}, statistical error) satellites
  brighter than $\MV{}=0$ within \Rout{} of the Sun.  As a result of our use of
  new data and better simulations, and a more robust statistical method,
  we infer a much smaller population of satellites than reported in
  previous studies using earlier \SDSS{} data only; we also address an
  underestimation of the uncertainties in earlier work by accounting
  for stochastic effects.  We find that the inferred number of faint
  satellites depends only weakly on the assumed mass of the MW
  halo and we provide scaling relations to extend our results to
  different assumed halo masses and outer radii. We predict that half
  of our estimated total satellite population of the MW should
  be detected by the Large Synoptic Survey Telescope.
  The code implementing our estimation method is available online.\footnotemark
\end{abstract}

\begin{keywords}
Galaxy: halo---galaxies: dwarf---dark matter.
\end{keywords}

%%%%%%%%%%%%%%%%%%%%%%%%%%%%%%%%%%%%%%%%%%%%%%%%%%

%%%%%%%%%%%%%%%%% BODY OF PAPER %%%%%%%%%%%%%%%%%%
\footnotetext{This is available from \citet{newton_mw_2018}.}
\section{Introduction}
\label{sec:Introduction}

Proposed in the 1980s \citep[e.g.][]{peebles_large-scale_1982,
davis_evolution_1985}, the $\Lambda$ cold dark matter (\LCDM{}) model has
proved remarkably successful at predicting numerous observable properties of the
Universe and their evolution over time; as a result, it has become the
`standard model' of cosmology \citep[see][for recent reviews]{frenk_dark_2012,
weinberg_cold_2015}. Hierarchical structure formation is fundamental to this
model, which predicts that dark matter~(DM) haloes form by mergers of smaller
haloes and smooth mass accretion. Merged (sub)haloes that are not completely
disrupted are detectable today as satellite galaxies and, potentially, as
non-luminous substructures.

The Milky Way (MW) halo and its associated satellite galaxies offer an ideal
environment in which to probe hierarchical growth which, in turn, can be used
to constrain the faint end of galaxy formation and the properties of the DM.
However, the current census of MW satellite galaxies is highly incomplete. The
most recent surveys---such as the Sloan Digital Sky Survey
\citep[\SDSS{};][]{alam_eleventh_2015} and the Dark Energy Survey
\citep[\DES{};][]{bechtol_eight_2015,drlica-wagner_eight_2015}---do not cover
the entirety of the sky and are also subject to detectability limits that
depend on the surface brightness of, and distance to the satellite galaxies.
The goal of this paper is to overcome some of these limitations and, using
theoretical priors based on cosmological simulations of MW-like haloes, to
estimate the expected total number of MW satellite galaxies.

In the 1990s, DM-only CDM simulations showed that many more subhaloes survive
within MW-like haloes than there are visible satellites orbiting the MW
\citep{klypin_where_1999,moore_dark_1999, springel_aquarius_2008}. This
disparity is often referred to as the `missing satellites problem for cold
dark matter.' This rather unfortunate nomenclature is very misleading if, as
is common usage, the word `satellite' is taken to mean a visible galaxy:
DM-only simulations have, of course, nothing to say about visible galaxies.
Simple processes, at the heart of galaxy formation theory, such as the 
reionization of hydrogen in the early universe and supernovae feedback, make it
impossible for visible galaxies to form in the vast majority of CDM haloes.
Such processes were first discussed and calculated in this context using
semi-analytic techniques with different approximations in the early 2000s
\citep{bullock_reionization_2000,benson_effects_2002,benson_effects_2002-1,
somerville_can_2002}. For example, \citet{benson_effects_2002} showed how the
abundance and stellar content of dwarf galaxies are driven by reionization and
supernovae feedback. Their model produced an excellent match to the luminosity
function of the ($11$ `classical'---the only known at the time) satellites of
the MW and predicted that the MW halo should host a large population of fainter
satellites. Just such a population was discovered several years later in the
\SDSS{} \citep[][and references therein]{koposov_luminosity_2008}.

The early semi-analytic results have been confirmed using full hydrodynamic
simulations \citep[e.g.][]{okamoto_effects_2005,maccio_concentration_2007}. For
example, the most recent such simulations have confirmed that below a certain
halo mass, typically \Msun[{\sim}10^{10}], dwarf galaxy formation is strongly
suppressed, and that the majority of haloes with masses \Msun[{\lesssim}10^9],
should not host a luminous component (stellar mass greater than \Msun[10^4])
\citep{shen_baryon_2014,sawala_bent_2015,sawala_chosen_2016,
wheeler_sweating_2015}.

In recent years, alternatives to CDM have elicited considerable
interest. Some of these, such as Warm Dark Matter
\citep[WDM,][]{avila-reese_formation_2001, bode_halo_2001}, models with
interactions besides gravity between DM particles and photons or neutrinos
\citep{boehm_using_2014} and axionic DM \citep{marsh_axion_2016},
predict a cut-off in the primordial matter power spectrum on
astrophysically relevant scales, which would suppress the formation of
small galaxies \citep{bode_halo_2001,polisensky_constraints_2011,lovell_haloes_2012,
  schewtschenko_dark_2015}. The abundance of the faintest galaxies
can thus, in principle, reveal or rule out the presence of a
power spectrum cut-off. By requiring that WDM models should produce at
least enough substructures to match the observed Galactic satellite
count, constraints on the mass and properties of the DM particle can
be derived
\citep{maccio_how_2010,lovell_properties_2014,kennedy_constraining_2014,schneider_astrophysical_2016,bose_substructure_2017,lovell_properties_2017}.

Past and current surveys have now discovered a plethora of satellites
around the MW, with the count currently standing at $56$: $11$
classical satellites, $17$ discovered in each of the SDSS and DES
surveys, and $11$ found in other surveys. Despite this relatively large
number of known satellites, current estimates suggest that there could
be at least a factor of $3\hbox{--}5$ times more still waiting to be
discovered
\citep{koposov_luminosity_2008,tollerud_hundreds_2008,hargis_too_2014}. These
estimates were made prior to the DES and are based only on SDSS
data. These predictions start from an assumed radial profile for the
distribution of Galactic satellites: either that it follows the DM
density profile---as in \citet{koposov_luminosity_2008}, which is not
a good assumption---or that it follows the subhalo number density
profile (as in the other studies cited above). Then, for each observed
satellite, they calculate the number of satellites in the entire
fiducial volume that must be present in order to have, on average, one
object with the corresponding properties within the survey volume.

This paper improves upon previous estimates of the Galactic satellite
count in three major ways. First, while previous studies were based
on SDSS data alone, our result makes use of the combined SDSS and DES
data, which together cover an area equivalent to nearly half
of the sky. Secondly, to properly account for stochastic effects, we
introduce a new Bayesian approach for estimating the total satellite
count. Stochastic effects---which we find to be the leading cause
of uncertainty---have been overlooked in previous studies, resulting
in a significant underestimation of their errors. Finally, we make use
of a set of five high-resolution simulated host haloes---taken from
the \Aquarius{} project \citep{springel_aquarius_2008}---to
characterize uncertainties arising from host-to-host variation. In
2016 December, \citet{jethwa_upper_2018} presented a Bayesian estimate
of the total number of Galactic satellites. Their result is the
outcome of applying abundance matching to the SDSS observations and,
while it properly accounts for stochastic effects, it depends on
more and uncertain assumptions (mostly related to abundance matching)
than the result presented here.

We organize this paper as follows. \secref{sec:Data} introduces the
observational data set used in this analysis and \secref{sec:Methods}
describes, tests, and compares our Bayesian technique with previous
works. We present our main results in \secref{sec:Results}, detailing
their sensitivity to the assumed MW halo mass and the radial
dependence of the satellite count. \secref{sec:Discussion} discusses
the implications of our results and considers some of the limitations
of our method. We present concluding remarks in
\secref{sec:Conclusions}.

\section{Observational Data}
\label{sec:Data}
Very few of the current set of MW satellites were known prior to the
start of the 21st century. Discoveries made after this time, using a
multitude of techniques, together with data from \SDSS{} data release
2 (\DR{2}) and the Two Micron All-Sky Survey (\twomass{})---before a
major advance with \SDSSDR{5} \citep{adelman-mccarthy_fifth_2007}---brought
the total to 23 dwarf galaxies.  Since then, the \SDSS{}
survey area has nearly doubled and \DES{} is now electronically
available. Combining the two surveys produces a sky coverage area of
$47$~per cent, with \SDSS{} and \DES{} contributing $14\, 555$ and $5000$
square degrees, respectively. An analysis of \DES{} data added a
further 17 dwarf galaxies to the running total
\citep{bechtol_eight_2015,drlica-wagner_eight_2015,kim_heros_2015,koposov_beasts_2015},
which, together with other discoveries, brings the total number of
dwarf galaxies, as of 2018 February, to $56$. These are listed in
\tabrefs{tab:obs_sat_table}{tab:sec_sat_table} of \appref{app:Known_satellites}.

These discoveries resulted from the use of advanced search algorithms
that comb through survey data and identify overdensities of stars
which could signal the presence of a faint dwarf galaxy. For example,
the \SDSS{} has been analysed with two such search algorithms,
by \citet{koposov_luminosity_2008} and \citet{walsh_invisibles_2009},
to find that both techniques recover the same number of dwarf
galaxies---although the latter is sensitive to fainter objects. Each algorithm
has a response function that---among other factors such as the
survey surface brightness limits---is dependent on the absolute
magnitude of the objects being searched for. Assuming isotropy, the
number of observed satellites per unit magnitude, ${\d{N_{\rm
      sat}}}/{\d{\MV{}}}$, is given by

\begin{equation}
	\frac{\d{N_{\rm sat}}}{\d{\MV{}}}{=}\int_0^{\infty} \int_0^{\infty} \Omega r^2 \frac{\d{^3N_{\rm sat}}}{\d{r}\;\d{\MV{}} \;\d{r_{\rm sat}}}  \; \epsilon(r,\MV{},r_{\rm sat}) \d{r} \d{r_{\rm sat}} \;,
	\label{eq:number_observed}
\end{equation}
where the first integral is over the survey volume, with $\Omega$ the
survey solid angle and $r$ the radial distance from the Sun. The
second integral is over the satellite size, $r_{\rm sat}$; $N$ is the
distribution of satellites as a function of radial distance from the Sun,
absolute magnitude, \MV{}, and size, $r_{\rm sat}$. The last term, $\epsilon$,
denotes the efficiency of the search algorithm for identifying a satellite
of magnitude, \MV{}, and size, $r_{\rm sat}$, at distance, $r$,
averaged over the survey's sky-footprint.
At fixed absolute magnitude, most of the satellites detected in the \SDSS{}
have similar sizes and the detection efficiency, $\epsilon$,
is approximately equal for all objects
\citep{koposov_luminosity_2008,walsh_invisibles_2009}. Thus, for
the observed satellites, the dependence on $r_{\rm sat}$ in
\eqnref{eq:number_observed} can be approximated as a dependence on
\MV{} alone.

The detection efficiency, $\epsilon$, at fixed \MV{}, is a function of the
radial distance and shows a rapid transition with radius from a $100$~per cent
to a $0$~per cent chance of detection. We may therefore define an equivalent
effective detection volume such that, on average, this effective volume
includes the same number of satellites of magnitude \MV{} as predicted by
\eqnref{eq:number_observed}. The effective radius,
$\Reff{} \left( \MV{} \right)$, corresponding to this effective detection
volume, is computed by solving the equation,
\begin{equation}
	\frac{\d{N_{\rm sat}}}{\d{\MV{}}}{=}\int_0^{\Reff{} \left( \MV{} \right)} \Omega r^2 \d{r} \frac{\d{^2N_{\rm sat}}}{\d{r}\;\d{\MV{}}} \;,
	\label{eq:number_observed_simplified}
\end{equation}
where the left-hand term is given by \eqnref{eq:number_observed} and
\Reff{} appears as the upper limit of the integral. The value of \Reff{}
depends on both the radial dependence of $\epsilon$ and on the radial
distribution of satellites. As long as the radial distribution of satellites is
nearly constant in the interval where the detection efficiency drops from $100$
to $0$~per cent, \Reff{} can be approximated as the radius at which the
detection efficiency is $50$~per cent, which is the value that we use in the
rest of this paper. This approximation is reasonable as $\epsilon$ decreases
from $1$ to $0$ over a narrow radial range
\citep[e.g. see fig.~$15$ in][]{walsh_invisibles_2009}.
Making another choice for the effective radius, such as $\epsilon = 0.9$ (as
used in \citealt{hargis_too_2014}), would underestimate the effective volume
and thus overestimate the inferred satellite count. Both
\citet{koposov_luminosity_2008} and \citet{walsh_invisibles_2009}
show that, to good approximation, the effective detection radius, which
corresponds to $\epsilon = 0.5$, is given by
\begin{equation}
	\Reff{} \left( \MV{} \right){=}10^{\left( -a^*\MV{} - b^* \right)}~\Mpc{} \;,
	\label{eq:R_eff}
\end{equation}
where $a^*$ and $b^*$ are fitting parameters associated with the
search algorithm response function. These values are provided in
\tabref{tab:Reff_params} for different algorithms.

\begin{table}
	\centering
	\caption{The parameters of \eqnref{eq:R_eff} quantifying the dependence on
	absolute $V$-band magnitude of the effective radius in the \SDSS{} and
	\DES{} surveys. The \citet{koposov_luminosity_2008} parameters are taken
	from fits by \citet{walsh_invisibles_2009}.}
	\label{tab:Reff_params}
	\begin{tabular}{lccc} % four columns, alignment for each
		\hline
		Survey & Algorithm & $a^*$ & $b^*$\\
		\hline
		\multirow{2}{*}{\SDSS{}} & \citet[K08]{koposov_luminosity_2008} & 0.205 & 1.72\\
		& \citet[W09]{walsh_invisibles_2009} & 0.187 & 1.58\\
		\DES{} & \citet[J16]{jethwa_magellanic_2016} & 0.228 & 1.45\\
		\hline
	\end{tabular}
\end{table}

\begin{figure}
	\vspace{-5pt}
	\includegraphics[width=\columnwidth]{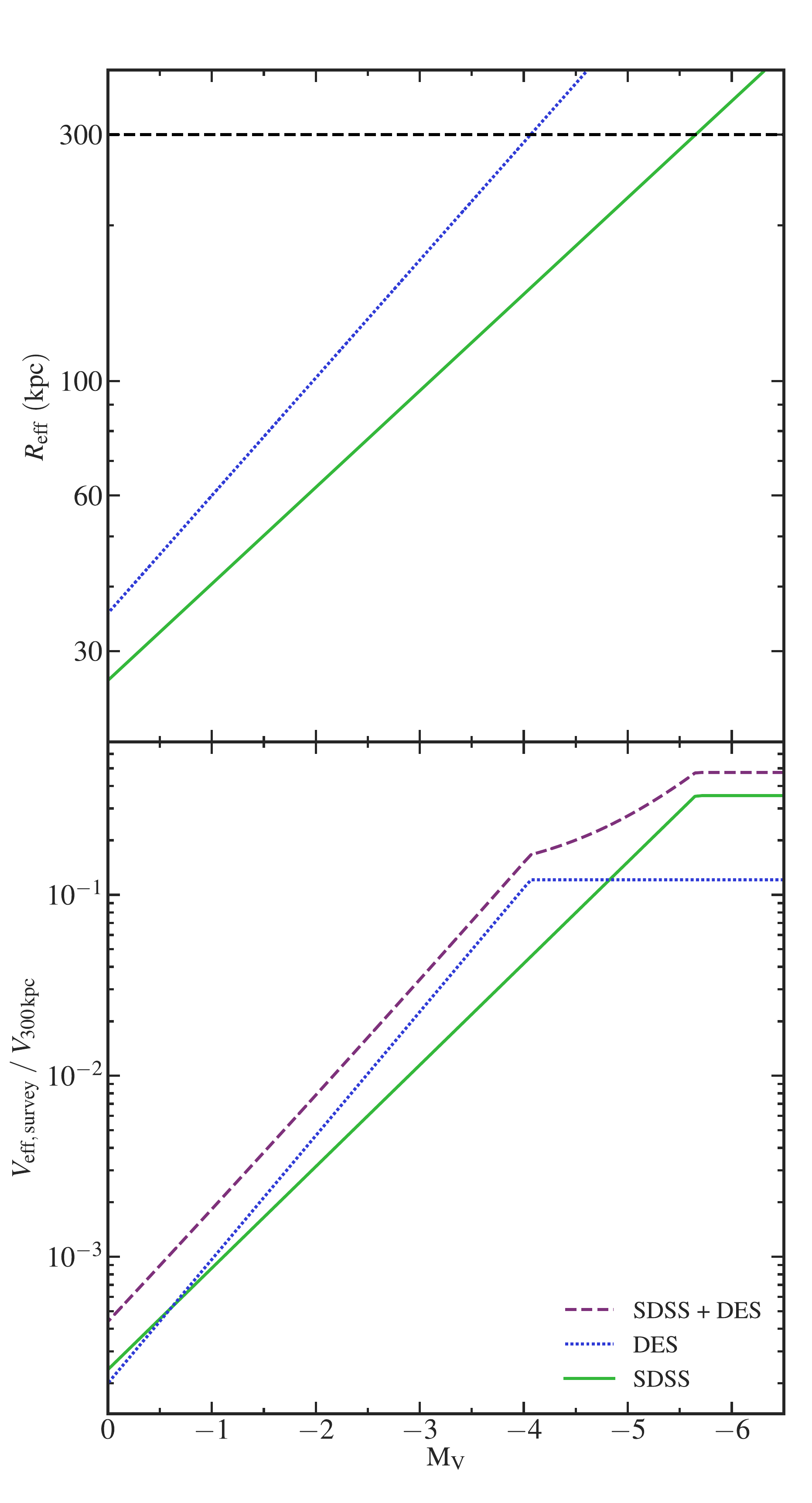}
	\vspace{-17pt}
	\caption{\textit{Upper panel:} the effective detection radius,
          $\Reff{}$, of satellites as a function of absolute
          magnitude, \MV{}, for the \SDSS{} and \DES{} surveys. The
          horizontal dashed line indicates our fiducial choice of
          outer radius, $\router{}{=}\Rout{}$, for the MW satellite
          population. \textit{Bottom panel:} the ratio of the
          effective volume surveyed by \SDSS{} and \DES{}, as a
          function of \MV{}, to the volume enclosed within \Rout{}. The
          dashed line shows the combined \SDSS{} plus \DES{} effective
          volumes. The two panels show the response functions of the
          \citetalias{walsh_invisibles_2009} and
          \citetalias{jethwa_magellanic_2016} search algorithms, which are
          given in \tabref{tab:Reff_params}.}
	\label{fig:Reffs}
	\vspace{-10pt}
\end{figure}

The dependence of the effective radius on absolute $V$-band magnitude
for the \SDSS{} and \DES{} surveys is shown in the upper panel of
\figref{fig:Reffs}. For clarity, in the case of the \SDSS{} we show only
the \citet{walsh_invisibles_2009} response function. For \DES{} we
give the \citet{jethwa_magellanic_2016} response function that was
shown to give a good match to the actual detections. This is equal
to the \citet{koposov_luminosity_2008} response function as fitted by
\citetalias{tollerud_hundreds_2008}, but shifted
to account for the additional depth of the \DES{} compared to
\SDSS{}; however, this response function has not been verified at the same level
of in-depth analysis as in e.g.~\citet{walsh_invisibles_2009}. The figure
shows that for the same absolute magnitude, \DES{} is deeper and thus can detect
satellites out to greater distances than \SDSS{}. All bright dwarfs,
i.e. $\MV{}<-5.5$ for \SDSS{} and $\MV{}<-4.0$ for \DES{}, that are within the
survey footprint and within our fiducial choice of outer radius,
$\router{}{=}\Rout{}$, should have been detected within their respective surveys.
Thus, the surveys may be considered `complete'---for the purposes of this
analysis---at the absolute magnitudes at which \Reff{} is greater
than \Rout{}. Fainter objects can be detected only if they are
closer than \Rout{} from the observer, with the faintest, $M_{\rm V}{=}0$,
dwarfs being detected only if they are within ${\sim}\kpc[30]$ of the Sun.

To obtain a more informative perspective on the survey completeness,
the bottom panel of \figref{fig:Reffs} shows the ratio between the
effective volume of each survey and the total volume enclosed within
our fiducial radius of \Rout{}. Even when combining the \SDSS{} and
\DES{} footprints, the observations cover only ${\sim}10$~per cent of the
fiducial volume at $\MV{}{=}-4$ and less than $0.1$~per cent of the same volume
at $\MV{}{=}0$.

\section{Methodology}
\label{sec:Methods}
We require two key ingredients to estimate the total population of satellite
galaxies from a given survey of the MW. First, we need a prior for the radial
distribution of satellites. For this we take the radial number density of
subhaloes in simulations of MW analogues from the \Aquarius{} project, which,
when subhaloes are selected by \vpeak{}---the highest maximum circular
velocity achieved in the subhalo's history---is the same as the radial
distribution of luminous satellites in hydrodynamic simulations and that of
observed MW satellites (see \secref{sec:Tracer_population}). Secondly, we
introduce and test our Bayesian framework used to infer the total number of
satellites (\secref{sec:Bayes_frame}). The need for a new methodology
is motivated by several shortcomings of previous approaches, which we
discuss in detail in \secref{sec:compare_methods}.

We assume that the classical satellites, i.e. those with $\MV{}\leq-8.8$, are
bright enough to have been observed by pre-\SDSS{} surveys and that
the observations are complete at these magnitudes (therefore ignoring the possible existence of obscured satellites in the Zone of Avoidance). As such, the
inferred luminosity function at the bright end will always match the
observations, in line with previous studies
\citep[e.g.][]{tollerud_hundreds_2008}. The inference method is only
applied to fainter satellites, that is, those with $\MV{}>-8.8$.

\subsection{Tracer population}
\label{sec:Tracer_population}
Any estimation of the total satellite count from incomplete
observations needs a prior for the radial number density of these
objects, which we estimate from N-body simulations. An ideal
simulation from which to extract a tracer population should have 
high enough resolution for the density profile to be well sampled, and
should also offer access to multiple realizations of MW-like haloes
to account for host-to-host variations.

The \Aquarius{} suite of simulations \citep{springel_aquarius_2008}
achieves this. It consists of a set of six \LCDM{} \mbox{DM-only} \mbox{N-body}
simulations of isolated MW-like haloes which were run using the
\GadgetIII{} code and were labelled \mbox{Aq-A} to \mbox{Aq-F}. In this work we use
the `level 2' simulations (L2, with a particle mass of
${\sim}\Msun[10^4]$), which corresponds to the highest
resolution level available across all of the \Aquarius{} haloes.
Details of these simulations are provided in \tabref{tab:Aq_sims}. The
\mbox{Aq-F} halo experienced a late-time merger, making it unsuitable as
representative of the MW halo; consequently, it is not used in this
analysis. The cosmological parameters assumed for these simulations
are derived from the \WMAP{} first-year data release
\citep{spergel_first-year_2003}: \WMAPonep{}.

\begin{table}
	\centering
	\caption{The DM particle mass, $m_{\rm p}$, softening length,
          $\epsilon$, and host halo mass, \Mvir{}, of the
          \Aquarius{} simulations used in this work. Here, \Mvir{}
          denotes the mass inside the radius,
          $R_{200}$, within which the mean density 
          equals $200$ times the critical density.} 
	\label{tab:Aq_sims}
	\begin{tabular}{lccc}
		\hline
		Simulation & $m_{\rm p}\,\left(\Msun{}\right)$ & $\epsilon\, \left(\pc\right)$ & $\Mvir{}\, \left( \Msun[10^{12}] \right)$\\
		\hline
		Aq-A1 & $1.712 \times 10^3$ & 20.5 & $1.839$\\
		Aq-A2 & $1.370 \times 10^4$ & 65.8 & $1.842$\\
		Aq-B2 & $6.447 \times 10^3$ & 65.8 & $0.819$\\
		Aq-C2 & $1.399 \times 10^4$ & 65.8 & $1.774$\\
		Aq-D2 & $1.397 \times 10^4$ & 65.8 & $1.774$\\
		Aq-E2 & $9.593 \times 10^3$ & 65.8 & $1.185$\\
		\hline
	\end{tabular}
\end{table}

Identifying subhaloes near the centre of simulated haloes using
configuration space halo finders like SUBFIND can be difficult
\citep{springel_aquarius_2008,onions_subhaloes_2012}. Subhalo finders are
affected by the resolution of the simulation to
which they are applied; these effects can be assessed by
comparing haloes which have been simulated at different resolution
levels. One of the haloes in the \Aquarius{} suite (\mbox{Aq-A}) was
simulated at extremely high-resolution (`Level~1' or L1, with particle
mass of \Msun[{\sim}10^3]). Even though the resolution of L2
is still very high, the abundance of subhaloes that are relevant to
our analysis is suppressed relative to that at L1, particularly in the
inner regions of the halo. The difference between the two levels is
comparable to that seen across all other L2 profiles. We can, however,
correct for these resolution effects in a relatively straightforward
manner, by using the Durham semi-analytic model
\Galform{} \citep{lacey_unified_2016,simha_modelling_2017} to populate the
haloes and subhaloes in the \Aquarius{} simulations with galaxies and
track their orbital evolution even after its halo is no longer resolved
(the so-called `orphan' galaxies).  The detailed scheme we used and a
comparison of the subhalo samples, both before and after application
of \Galform{}, are given in \appref{app:Resolution_effects}.

\begin{figure}
	\includegraphics[width=\columnwidth]{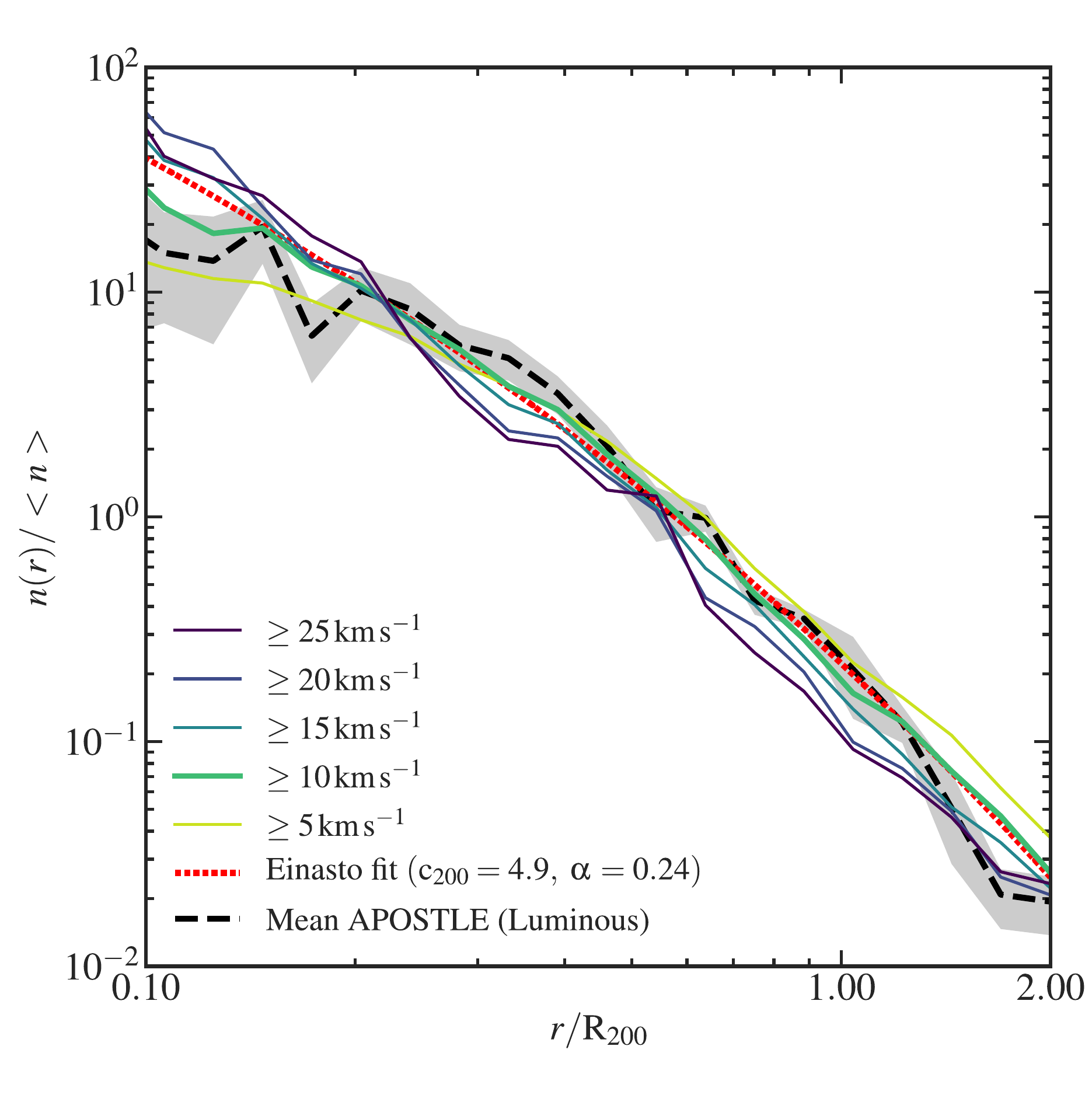}
	\vspace{-17pt}
	\caption{The radial number density of fiducial subhaloes normalized to
          the mean density within \Rvir{}. The thin solid lines show
          the distributions for subhaloes with different \vpeak{} cuts
          averaged over the five \Aquarius{} haloes. The thick dashed
          line and associated shaded region show the radial
          distribution of \textit{luminous} satellites and its
          associated $68$~per cent scatter obtained using eight haloes from
          the \Apostle{} high-resolution hydrodynamic simulations. The
          thick dotted line shows the best-fitting Einasto profile. For
          ease of comparison the profile with our chosen selection
          criterion of $\vpeak{}{\geq}\kms[10]$ is provided as a thick
          solid line.  }
	\label{fig:aq_ap-lum_comparison}
	\vspace{-10pt}
\end{figure}

A further factor that needs to be taken into account is the possible
destruction of satellite galaxies by tidal interactions with the
central galaxy in the halo. This effect has been calculated by
\citet[fig.~4,~upper~panel]{sawala_shaken_2017} using the \Apostle{}
hydrodynamic simulations that show that up to $40$~per cent of satellites in
the inner ${\sim}\kpc[30]$ can be destroyed, although overall the
fraction destroyed is much smaller \citep[see also][]
{donghia_substructure_2010,errani_effect_2017,garrison-kimmel_not_2017}.
For our purposes this difference, which changes the radial subhalo
distribution, is fairly important but it has the opposite effect to
the omission of orphan galaxies and, as we discuss below, the two
effects partially cancel out.  To correct for these baryonic effects, we
downsample the $z{=}0$ \Aquarius{} subhaloes according to the value of
the radius-dependent depletion rate derived by
\citet{sawala_shaken_2017}.\footnote{There is an error in the values of the
   fitting parameters quoted by \citet{sawala_shaken_2017}; see
   \appref{app:Baryonic_effects} for further details and the correct values of
   the parameters.}
The radial dependence of the depletion
factor and further details about this procedure are given in
\appref{app:Baryonic_effects}. We refer to this final population,
which incorporates `orphan galaxies' and baryonic effects, as our
fiducial tracer population. Unless otherwise stated we use this
subhalo population throughout the rest of this paper.

We apply a selection cut to the fiducial \Aquarius{} subhalo
populations on the basis of their \vpeak{} values, under the
expectation that this will provide a stronger correlation with the
likelihood of a galaxy forming within the subhalo
\citep{sawala_chosen_2016} than, for example, selecting by present-day
maximum circular velocity or present-day mass
\citep{libeskind_distribution_2005,wang_spatial_2013}. This correlation has
been shown to hold in the \LCDM{} model, which is one of the priors in our
analysis. In \figref{fig:aq_ap-lum_comparison} we show the radial number density
of subhaloes normalized by the mean subhalo density within \Rvir{}. This
is used to assess the appropriateness of applying a \vpeak{}
selection, and to determine the \vpeak{} value down to which the
profiles are consistent. We compare this against the radial
distribution of luminous satellites selected from a set of high-resolution
hydrodynamic simulations from the \Apostle{} project
\citep{fattahi_apostle_2016,sawala_apostle_2016}. This is a suite of
$12$ cosmological zoom resimulations of Local Group-like regions run
with the \GadgetIII{} code and \Eagle{} subgrid physics models
\citep{schaye_eagle_2015,crain_eagle_2015}. Of these, $4$
regions---which contain $8$ MW and M31 analogues
-- were re-run at much higher resolution and are used here. The
\Apostle{} data are not used beyond the provision of this reference
profile as the simulation is unable to resolve ultrafaint luminous
satellites at the magnitudes we are considering here.

\figref{fig:aq_ap-lum_comparison} shows that the radial profile of
subhaloes is largely independent of the value of \vpeak{}, except for
values below \kms[10], where resolution effects come into play. Most
importantly, we find that the profiles of samples selected with
thresholds above this value are in good agreement with the
profile of the luminous \Apostle{} satellites, and that of observed MW satellites (see \secref{sec:tracer_MW_sat_comparison}), making this a good
choice to model the radial distribution of satellites. We therefore
only consider subhaloes with $\vpeak{}{\geq}\kms[10]$ in the rest of
our analysis.

\subsubsection{Rescaling the \Aquarius{} haloes to a fiducial MW halo mass}
\label{sec:mass_rescaling}
We would like to assess if the calculation of the total satellite
count is sensitive to the mass of the MW halo. This is important in
view of the large uncertainties in current estimates of the MW halo
mass, with values typically in the range
\Msun[\left(0.5-2.0\right)\times10^{12}]
\citep[e.g.][]{piffl_rave_2014, cautun_milky_2014,
  wang_estimating_2015}. To do this, we rescale the Aquarius haloes to
a fiducial MW halo mass, $M_{\rm MW, target}$, and apply our Bayesian
method to these rescaled haloes. When expressed as a function of
rescaled radial distances, $r\, /\, \Rvir{}$, the radial number density of
subhaloes is largely independent of host mass
\citep{springel_aquarius_2008,han_unified_2016,hellwing_copernicus_2016}. Thus,
we can rescale the original Aquarius haloes to different target masses
by multiplying the radial distance of each subhalo by the ratio
$R_{\rm 200,\, target}\, /\, R_{\rm 200,\, original}$. Unless specified
otherwise, the results presented in this paper are calculated for a
fiducial MW halo mass, $M_{\rm MW}{=}\Msun[1.0\times10^{12}]$. The
variation of these results with MW halo mass is analysed in
\secref{sec:Halo_mass_influences}.

\subsubsection{Comparison to the MW satellite distribution}
\label{sec:tracer_MW_sat_comparison}
\begin{figure}
	\includegraphics[width=\columnwidth]{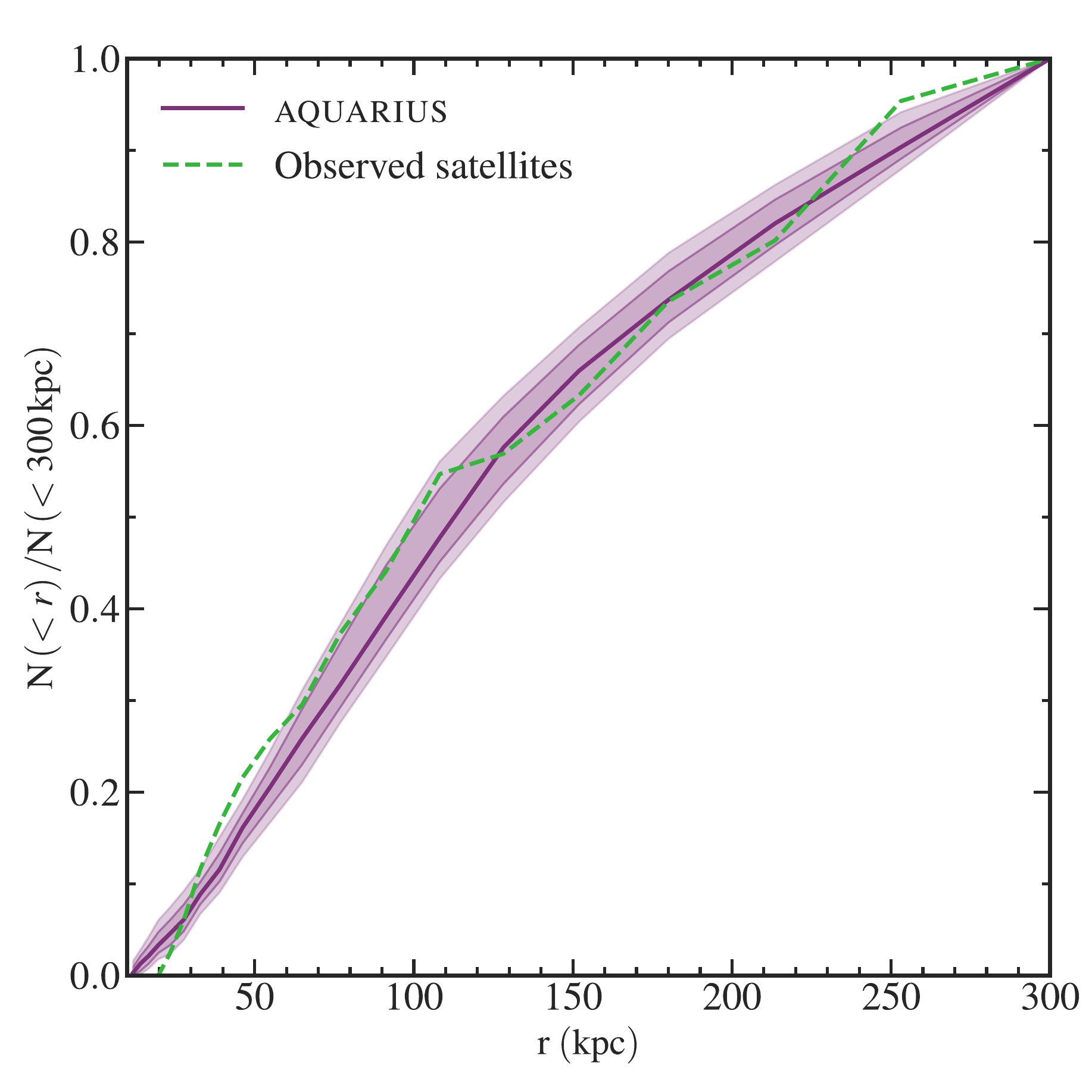}
	\vspace{-17pt}
	\caption{Comparison of the radial distribution of observed MW
	satellites (dashed line) with that of \vpeak{}-selected subhaloes from the
	five \Aquarius{} haloes (solid line) rescaled to a host halo mass of
	\Msun[1.0\times10^{12}]. The sample of observed satellites was corrected for
	survey radial incompleteness (see the text) and consists of the classical,
	\SDSS{}, and \DES{} satellites. We further accounted for the possibility
	that many of the \DES{} satellites may have fallen in with the
	\LMC{} by using the probabilities of association with the MW given by
	\citet{jethwa_magellanic_2016}. The dark and light shaded regions represent
	the \CL{68\ {\rm per\ cent}} and \CL{95\ {\rm per\ cent}}~(statistical
	error) bootstrapped error regions for the
	\vpeak{}-selected subhalo distribution, respectively.
    }
	\label{fig:radial_distribution_MW}
	\vspace{-10pt}
\end{figure}

A further test of the appropriateness of a particular choice of tracer
population can be obtained by comparing its radial distribution with that of
the observed MW satellites. When calculating the latter,
we need to correct for the radial incompleteness in the surveys: faint
satellites can be detected only at small radial distances which, if unaccounted
for, leads to a biased, more centrally concentrated satellite distribution. This
radial profile, corrected for radial incompleteness, is given by
\begin{equation}
	\frac{\d N \left( r \right)}{\d r} = \frac{ \sum_{i} P_{\rm MW,\;i} ~\delta\left(r_i-r \right)}{ \sum_{i} P_{\rm MW,\;i} ~ \epsilon\left(r,M_{\rm V,\; i} \right)}
    \label{eq:corrected_obs_dist} \;,
\end{equation}
where the sum is over all the observed classical, \SDSS{} and \DES{} satellites,
$r_i$ and  $M_{\rm V,\; i}$ are the position and absolute magnitude of the
$i$-th satellite, and $\delta\left(r_i-r \right)$ is the Dirac delta function.
The quantity, $P_{\rm MW,\;i}$, denotes the probability that a satellite is
associated with the MW, which we take to be $1$ for all objects except the
\DES{} satellites. Many of these are likely to have fallen in as satellites of
the Large Magellanic Cloud~(\LMC{}) and, being at first infall, are still
concentrated near the position of the \LMC{} which is adjacent to the relatively
small region surveyed by the \DES{}. For these objects we use the probabilities
of association given by \citet{jethwa_magellanic_2016}; we discuss this point in
greater detail in \secref{sec:Separate_estimates} below. The quantity, $\epsilon$,
is the detection efficiency (see \secref{sec:Data}) at distance, $r$, for
satellites of magnitude, \MV{}, and accounts for radial incompleteness. The
denominator of \eqnref{eq:corrected_obs_dist} is maximal for small $r$ values,
where all observed satellites have $100$~per cent detection efficiency, and
decreases at large $r$.

\figref{fig:radial_distribution_MW} shows that \vpeak{}-selected subhaloes have the same radial distribution as the observed MW satellites, as predicted by theoretical arguments \citep{libeskind_distribution_2005}. 
This comparison demonstrates the validity of our fiducial choice
for the radial distribution of satellites. The subhalo distribution given in \figref{fig:radial_distribution_MW} corresponds to a MW halo mass of
\Msun[1.0\times10^{12}] and using a slightly lower value for the MW halo mass leads to an even better agreement between the two radial distributions.

We also used \eqnref{eq:corrected_obs_dist} to compute the model-independent
radial number density for three different observational subsamples: the
classical, \SDSS{}, and \DES{} satellites. We find good agreement between the
three subsamples (not shown), indicating that the data are consistent with the
radial distribution being independent of satellite brightness. This is
consistent with \figref{fig:aq_ap-lum_comparison}, where we find that the
radial profile of \vpeak{}-selected objects is largely independent of the value
of \vpeak{}.

\subsubsection{A fit to the radial profile of subhaloes}
\label{sec:radial_profile_fit}
In a later part of our analysis
(\secref{sec:Outer_radii_predictions}), we will make use of a
functional form for the radial profile of satellites in order to scale
our results to different MW halo masses or fiducial volumes. For this,
we fit an Einasto profile
\citep{einasto_jaan_construction_1965,navarro_inner_2004}\footnote{A fit to
the DM density profile of this form was first introduced in
\citet{navarro_inner_2004} but only referred to as the ``Einasto profile'' in
\citet{merritt_empirical_2006}.}
to the $\vpeak{}{\geq}\kms[10]$ curve shown in
\figref{fig:aq_ap-lum_comparison}.  The Einasto profile---or the very
similar NFW profile \citep{navarro_simulations_1995,navarro_structure_1996,
navarro_universal_1997}---provides
a good description of the radial number density of
substructures
\citep{sales_satellite_2007,kuhlen_via_2008,springel_aquarius_2008,han_unified_2016}.
We can parametrize the Einasto profile in terms of a shape parameter,
$\alpha$, and the concentration, $\conc{}{=}\Rvir{}\, /\, r_{-2},$
with $r_{-2}$ the scale radius at which the logarithmic slope of the
profile is $-2$. Using the scaled radial distance,
$\chi{=}r\, /\, \Rvir{}$, the Einasto profile is given by
\begin{equation}
	\label{eq:num_dens_prof_fit}
	\frac{n \left(\chi\right)}{\left<{\rm n}\right>}{=}\frac{\alpha\conc{}^3}{3\left(\dfrac{\alpha}{2}\right)^{\frac{3}{\alpha}}\LowIncGammaFunc{\frac{3}{\alpha}}{\frac{2}{\alpha}\conc{}^\alpha}} \exp\left[-\dfrac{2}{\alpha}\left(\conc{}\chi\right)^\alpha\right] \;,
\end{equation}
where $\left<{\rm n}\right>$ is the mean number density within \Rvir{} and the lower incomplete Gamma function, $\gamma$, is defined as
\begin{equation}
	\label{eq:Low_Inc_Gamma_Func}
	\LowIncGammaFunc{s}{x}{=}\int_{0}^{x}t^{s-1}\exp\left(-t\right)\,\d{t} \;.
\end{equation}
We find that an Einasto profile with $\conc{}{=}\fitconc{}$ and
$\alpha{=}\fitalpha{}$ provides a good match to the radial number density
of subhaloes, as may be seen in \figref{fig:aq_ap-lum_comparison}.

\subsection{The Bayesian inference method}
\label{sec:Bayes_frame}
We are interested in calculating the probability distribution function
(PDF) of the total number of satellites, $N_{\rm tot}(<\MV{})$, if
a survey with effective volume, $V_{\rm eff}(\MV{})$, has detected
$N_{\rm obs}(<\MV{})$ satellites. Note that both the effective
volume and the number of satellites are functions of absolute
magnitude; however, for ease of readability, we drop the explicit
dependence on \MV{}. Within the Bayesian formalism, the posterior
probability of having a total of $N_{\rm tot}$ satellites given that
we observe $N_{\rm obs}$ objects within a volume, $V_{\rm eff}$, is
given by
\begin{equation}
	\label{eq:Bayes_theorem}
	P\left(N_{\rm tot}\left| N_{\rm obs},V_{\rm eff}\right. \right) = \frac{ P\left(N_{\rm obs}\left| N_{\rm tot},V_{\rm eff}\right. \right) P\left(N_{\rm tot}\right) } { P\left(N_{\rm obs},V_{\rm eff}\right) } \;,
\end{equation}
where $P\left(N_{\rm obs}\left| N_{\rm tot},V_{\rm eff}\right. \right)$ is the
likelihood of having $N_{\rm obs}$ objects within volume $V_{\rm eff}$ if there
is a total of $N_{\rm tot}$ satellites. For the prior,
$P\left(N_{\rm tot}\right)$, we take a flat distribution; the denominator is a
normalization factor. Thus, we have
\begin{equation}
	\label{eq:Bayes_theorem_2}
	P\left(N_{\rm tot}\left| N_{\rm obs},V_{\rm eff}\right. \right) \varpropto P\left(N_{\rm obs}\left| N_{\rm tot},V_{\rm eff}\right. \right) \;.
\end{equation}
The method needs two more ingredients: (1) a prior for the radial distribution of satellites, which we take as that of \Aquarius{} \vpeak{}-selected subhaloes, and (2) a sample of observed satellites, which we take as that of the \SDSS{} and \DES{} surveys. Thus, $N_{\rm tot}$ represents the inferred total number of MW satellites given these priors.

In practice, it is computationally prohibitive to evaluate the likelihood
function over the full parameter space so we use Approximate Bayesian
Computation~(ABC). ABC methods approximate the likelihood by selecting
model realizations that are consistent with the data. For our study, ABC is
an accurate way to estimate the likelihood function because (i) we compare
the realizations with the actual data rather than with summary statistics
and (ii) our data set consists of a discrete number of satellites and our
method selects realizations that exactly reproduce the observations.

The likelihood can be computed using a Monte Carlo method
applied to each \Aquarius{} halo. We start by selecting the satellite
tracer population---i.e. the DM subhaloes---within our fiducial MW
halo radius and organizing them into a randomly ordered
list. Then, for each observed satellite, we estimate the required
number of satellites of equal brightness such that there is only one
such object inside the effective survey volume corresponding to that
observed dwarf galaxy. Starting with the brightest observed
satellite, we pick random numbers, $N_{\rm rand}$, until we find that
only one of the top $N_{\rm rand}$ subhaloes is inside the
corresponding effective survey volume. The resulting $N_{\rm rand}$
value corresponds to one possible realization of the total count of
objects, $N_{\rm tot}(\MV{})$, of brightness equal to that of the
observed satellite. We then remove the top $N_{\rm rand}$ subhaloes
and repeat the same procedure for the next brightest observed
satellite.

We considered ordering the subhalo list according to their \vpeak{}
values, which is equivalent to ordering them from brightest to
faintest, assuming that \vpeak{} is a luminosity indicator. This ordering
would have the advantage of capturing correlations between the
luminosity of spatially close satellites as would happen in the case
of group accretion. For example, a massive satellite at first infall
is likely to bring with it other luminous galaxies
\citep{wang_spatial_2013,shao_alignments_2016}. In practice, we find
that the effects of any such correlations are insignificant compared
to the uncertainties introduced by host-to-host variability.

This Monte Carlo procedure generates one possible realization of the
dependence of the total number of satellites on absolute magnitude,
$N_{\rm tot}(<\MV{})$. To sample the full allowed space, the
procedure must be repeated many times, for different locations of the
survey volume, for different host haloes, and for new randomizations
of the subhalo list. The details of how we
achieve this are given in \secref{sec:Apply_bayes_frame}, together
with a more computationally efficient implementation of the Monte
Carlo algorithm just described.

Our Monte Carlo approach represents a discrete sampling of the effective
volume, $V_{\rm eff}$, which is a smooth function of \MV{}. While in
principle this may lead to biases, in practice there are enough observed
satellites to sample densely the range of absolute magnitudes of interest;
thus, any such effects are small, as may be seen in
\secref{sec:Validation}.

\subsubsection{Practical implementation}
\label{sec:Apply_bayes_frame}
For each \Aquarius{} halo, we position an observer \kpc[8] from the
halo centre at one of six vertices of an octahedron, and select a spherical
region of \Rout{} in radius centred on this point, similar to
\citet{tollerud_hundreds_2008}. All subhaloes within this region are
sorted randomly and assigned an index. We then
select a conical region with its apex at the observer position and its
opening angle corresponding to the sky coverage of the survey from
which the observational data are drawn. The maximum radial extent of
the conical region, \Reff{}, for an observed object of given magnitude
is calculated using \eqnref{eq:R_eff}.

Starting with the brightest object in the survey, of magnitude
$M_{\rm V,\, 1}$, we sequentially select subhaloes from our sorted list until
we identify one object within our mock survey volume. This sets the
lower bound for $N_{\rm tot}(< M_{\rm V,\, 1})$. To set the upper bound,
we continue down the sorted list of subhaloes until we find the
largest subhalo index which still corresponds to only one subhalo
inside the mock survey volume. Every choice between the lower and upper
bounds is equally consistent with the observation of one object of 
$M_{\rm V,\, 1}$ within the survey volume; we therefore randomly select
one number in this interval and remove this many subhaloes from
the beginning of our ordered list. We then consider the next brightest
object---of magnitude $M_{\rm V,\, 2}$---and repeat the above
procedure, using the updated list of subhaloes and
the new effective survey volume, $V_{\rm eff}(M_{\rm V,\, 2})$. We
continue this process down to the faintest observed satellites in the survey.

The procedure is repeated for $1000$ pointings evenly distributed
across the simulated sky, and for six observer locations, creating $6000$
realizations for each simulated halo. There are $5$ \Aquarius{} haloes so, in
total, we obtain $3\times10^4$ realizations that are used to estimate 
the median and $68$~per cent, $95$~per cent, and $98$~per cent uncertainties of
the complete satellite luminosity function.

\subsubsection{Validation}
\label{sec:Validation}
\begin{figure}
	\includegraphics[width=\columnwidth]{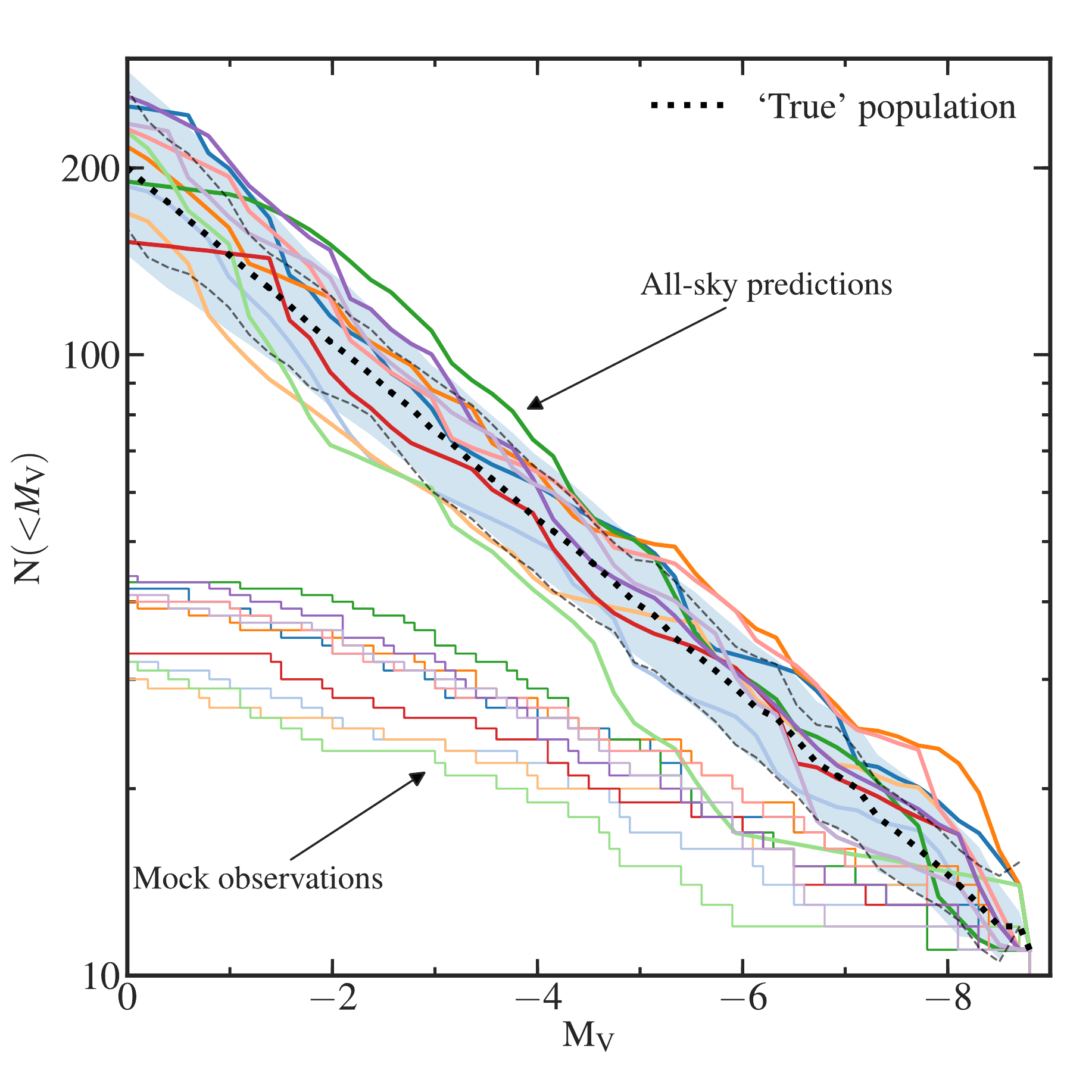}
	\vspace{-17pt}
	\caption{Tests of the Bayesian inference method using mock
          observations. The thick dotted line shows the input
          luminosity function used to create $100$ \SDSS{} mock
          observations. The luminosity functions of a sample of $10$ of these are
          shown as thin solid lines. Each of the $10$ mock
          observations was used, in turn, to predict a cumulative satellite
          luminosity function. The results are shown as thick solid
          lines. The shaded region represents the $68$~per cent uncertainty
          from one of the mock predictions, shifted to lie on top of the input
          luminosity function. The dashed lines bound the $68$~per cent
          confidence region over the medians of all $100$ mock predictions. }
	\label{fig:SDSS_mock_tests}
	\vspace{-10pt}
\end{figure}

In order to validate the Bayesian inference method, one of the authors
(ON) tested it on a set of $100$ mock \SDSS{} observations provided by
another (MC). The results of these tests, and a sample of $10$ of the mocks, are shown in
\figref{fig:SDSS_mock_tests}. The mock observations were generated
from a `blinded' luminosity function---indicated in the figure by the
thick dotted line---and were obtained
from the Aq-A1 halo distribution of subhaloes with $\vpeak{} \geq \kms[10]$ within \Rout{}.
The selected
subhaloes were then randomly assigned absolute magnitudes
according to the input luminosity function. Mock observations were
produced for $100$ random pointings of a conical region analogous to the
\SDSS{} volume within the halo, taking into account the effective
radius out to which satellites of different magnitudes could be identified. 
To model better the observations, mocks were generated using a radially
dependent detection efficiency: for a given magnitude, using \eqnref{eq:R_eff},
we calculated \Reff{}, which is the radius corresponding to a $50$~per cent
detection efficiency, and then assumed that the detection efficiency decreases
from $1$ to $0$ linearly in the radial range $[0.5,1.5]\, \Reff$. Satellites
found in regions where the detection efficiency is below unity were included in
the mocks using a probabilistic approach by comparing a random number between
$0$ and $1$ with the value of the detection efficiency.
The luminosity functions for a sample of $10$ of the $100$ resulting
mocks are shown as thin solid lines in
\figref{fig:SDSS_mock_tests}. Even though all the mocks survey the
same halo, we find a large spread in the number of observed
satellites.

Taking each mock survey data set in turn, we apply the Bayesian
inference method, producing $100$ estimates of the total satellite
luminosity function, $10$ of which are shown in \figref{fig:SDSS_mock_tests} as
thick solid lines. To assess the method fully, we also illustrate the
$68$~per cent uncertainty region, taken from one of the mocks and shifted
so that the centre of the region is aligned with the `true' luminosity
function. Most of the inferred satellite luminosity functions lie
inside the $68$~per cent uncertainty region, in line with statistical
expectations, thus demonstrating the success of the method at
reproducing the underlying true luminosity function. This uncertainty region,
taken from one mock, is comparable to the $68$~per cent confidence region
obtained from the medians of all $100$ mocks, which further demonstrates that
the method successfully estimates uncertainties. Note also that our inference
method assumes that the detection efficiency is a step function at \Reff{}, but
the mocks were generated using a radially varying detection efficiency. Thus,
this test also shows that assuming an effective detection radius is a good
approximation and does not bias the inferred total luminosity function.

\subsection{Comparison to previous inference methods}
\label{sec:compare_methods}
\begin{figure}
	\includegraphics[width=\columnwidth]{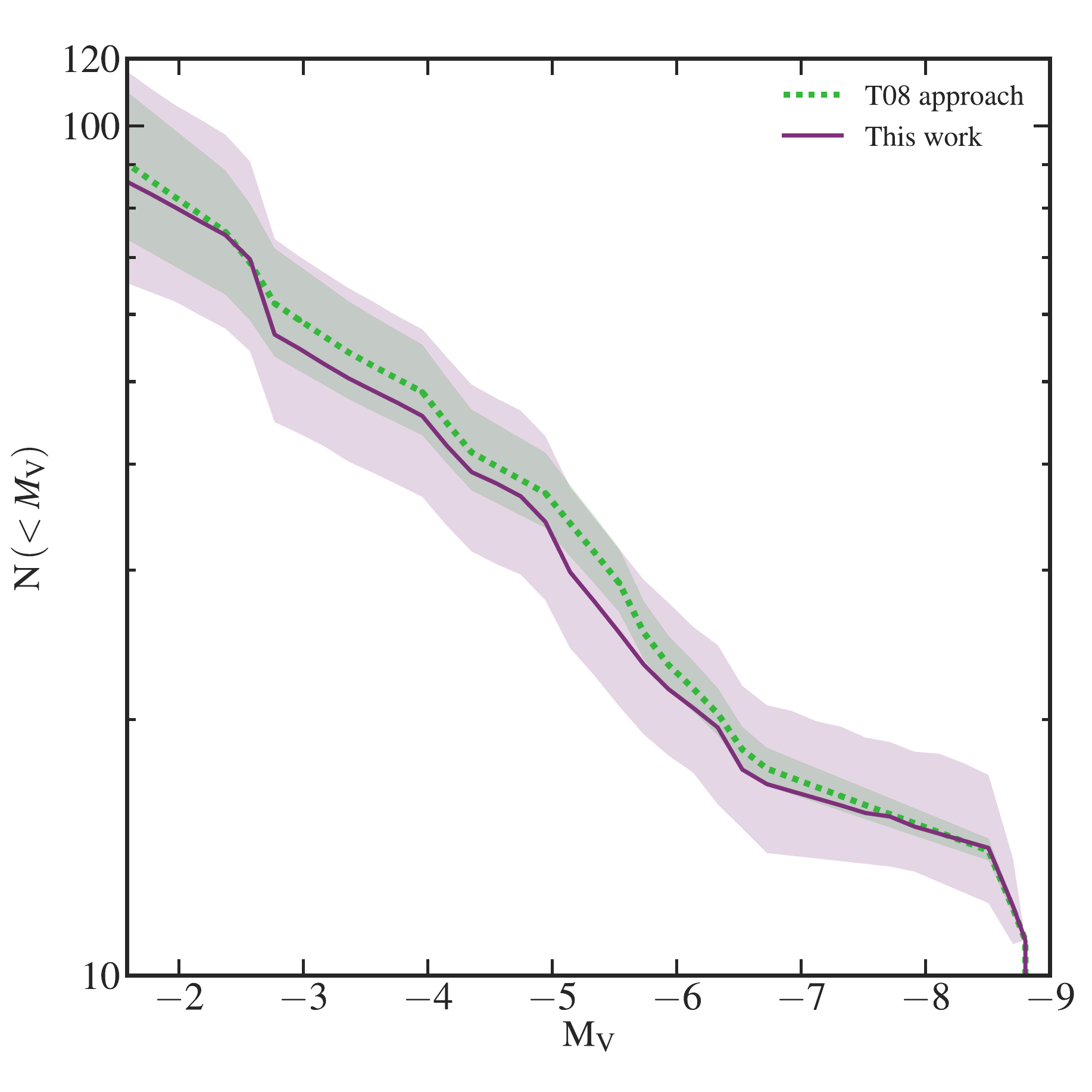}
	\vspace{-17pt}
	\caption{Comparison of two different inference methods for
          the total dwarf galaxy luminosity function: the
          \citet[T08]{tollerud_hundreds_2008} method and the Bayesian approach
          introduced here. Both methods were applied to the same
          data set, the \SDSS{}. The median estimate (solid line) and
          associated $68$~per cent uncertainties (shaded regions) for each
          method are shown. The \citetalias{tollerud_hundreds_2008}
          method does not account for stochastic effects,
          so it underpredicts the uncertainties.}
	\label{fig:aq_extrap_methods}
	\vspace{-10pt}
\end{figure}

As we discussed briefly in \secref{sec:Introduction}, the previous method used
for inferring the total satellite count has some drawbacks. The
\citet[][T08]{tollerud_hundreds_2008} method, which was also employed by
\citet{hargis_too_2014}, used a similar \vpeak{}-selected radial distribution of
subhaloes as us (although not accounting for unresolved subhaloes or baryonic
effects). However, the differences arise from the way in which these distributions
are used. The \citetalias{tollerud_hundreds_2008} method employs a completeness
volume, $V_{\rm comp},$ that is typically selected as the volume where the detection efficiency,
$\epsilon(\MV{})$, has a given non-zero threshold value, e.g.
$\epsilon(\MV{})=0.9$. Note that the \citetalias{tollerud_hundreds_2008}
completeness volume can be different from the effective volume used in our
Bayesian method. To obtain an unbiased estimate, only observed satellites within
that completeness volume, i.e. satellites with detection efficiencies above the
threshold value, should be used for inferring the total satellite count. The
\citetalias{tollerud_hundreds_2008} approach is based on
calculating, for each observed satellite, the fraction of \vpeak{}-selected
subhaloes inside the completeness survey volume associated
with that satellite. This fraction,
$\eta{=}N_{\rm sub}(<V_{\rm comp}) / N_{\rm max\;sub}$, is the ratio of the
number of subhaloes, $N_{\rm sub}(<V_{\rm comp})$, inside $V_{\rm comp}$ to the
total number of subhaloes, $N_{\rm max\;sub}$, inside the halo. Then, for the
$i$-th observed satellite, the fiducial halo volume contains
\begin{equation}
	\label{eq:T08_estimate}
	\frac{1}{\eta_i\epsilon_i} \;
\end{equation}
satellites of absolute magnitude, $\rm{M}_{{\rm V},\; i}$, with
$\epsilon_i$ the detection efficiency associated to the $i$-th observed satellite. 

\figref{fig:aq_extrap_methods} shows a comparison of the
\citetalias{tollerud_hundreds_2008} approach, discussed above,
with our Bayesian inference approach. These methods were applied to the
same \SDSSDR{9} data set using the \citet[W09]{walsh_invisibles_2009} 
completeness function (see \tabref{tab:Reff_params}) and the subhalo
distribution of a single simulated halo, Aq-A1, corrected for `orphan
galaxies' and baryonic effects. 
Here, when applying the \citetalias{tollerud_hundreds_2008} method, we choose a
completeness radius corresponding to $\epsilon(\MV{})=0.5$, which is equal to
the effective radius used by the Bayesian method, and only use observed
satellites with detection efficiencies, $\epsilon\geq0.5$. All the satellites
detected by the \citetalias{walsh_invisibles_2009} algorithm have
$\epsilon>0.5$ and thus pass this selection criterion.
The median estimates produced by the \citetalias{tollerud_hundreds_2008} and
Bayesian methods are similar. However, as we show in extensive tests detailed
in \appref{app:Testing_prev_methods}, where we apply the
\citetalias{tollerud_hundreds_2008} approach to mock observations similar to
those in \figref{fig:SDSS_mock_tests}, the \citetalias{tollerud_hundreds_2008}
method underestimates the uncertainties.

\begin{figure}
	\includegraphics[width=\columnwidth]{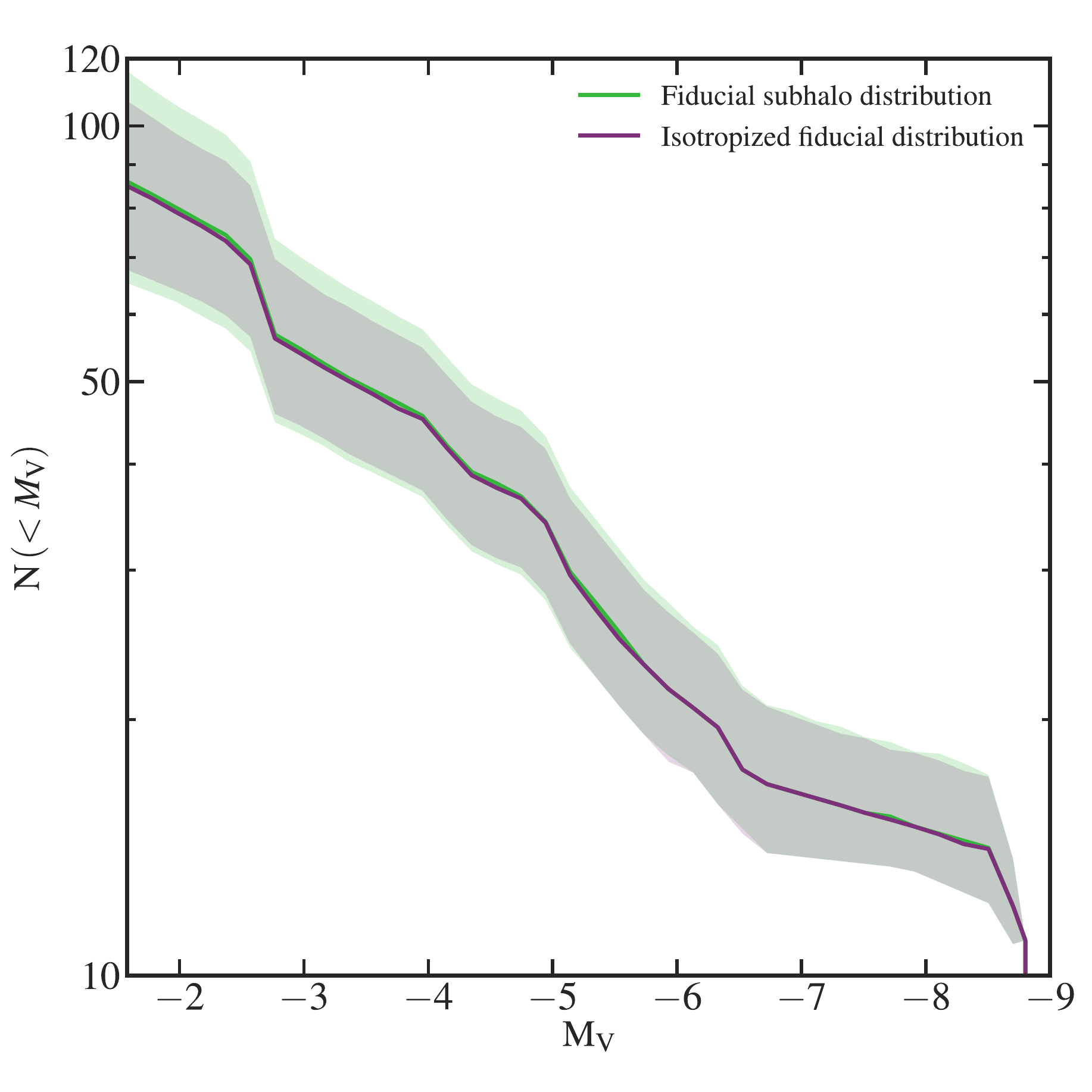}
	\vspace{-17pt}
	\caption{Comparison of the dominant sources of uncertainty in
          estimates of the total satellite luminosity function: the
          flattening of the subhalo distribution or the stochastic
          effects. The region labelled `fiducial subhalo distribution'
          corresponds to applying our method to the fiducial subhalo
          population of the simulated halo, Aq-A1. This estimate is
          affected by both the shape of the tracer distribution and
          stochastic effects. The region labelled `isotropized
          fiducial distribution' assumes the same radial distribution
          of subhaloes but with isotropized angular coordinates; this
          is affected only by stochastic effects. Both approaches
          have approximately the same median (solid line) and
          $68$~per cent scatter (shaded region). Thus, stochastic
          effects are a major source of uncertainty.  }
	\label{fig:SDSS_ido}
	\vspace{-10pt}
\end{figure}

\begin{figure*}
	\includegraphics[width=\textwidth]{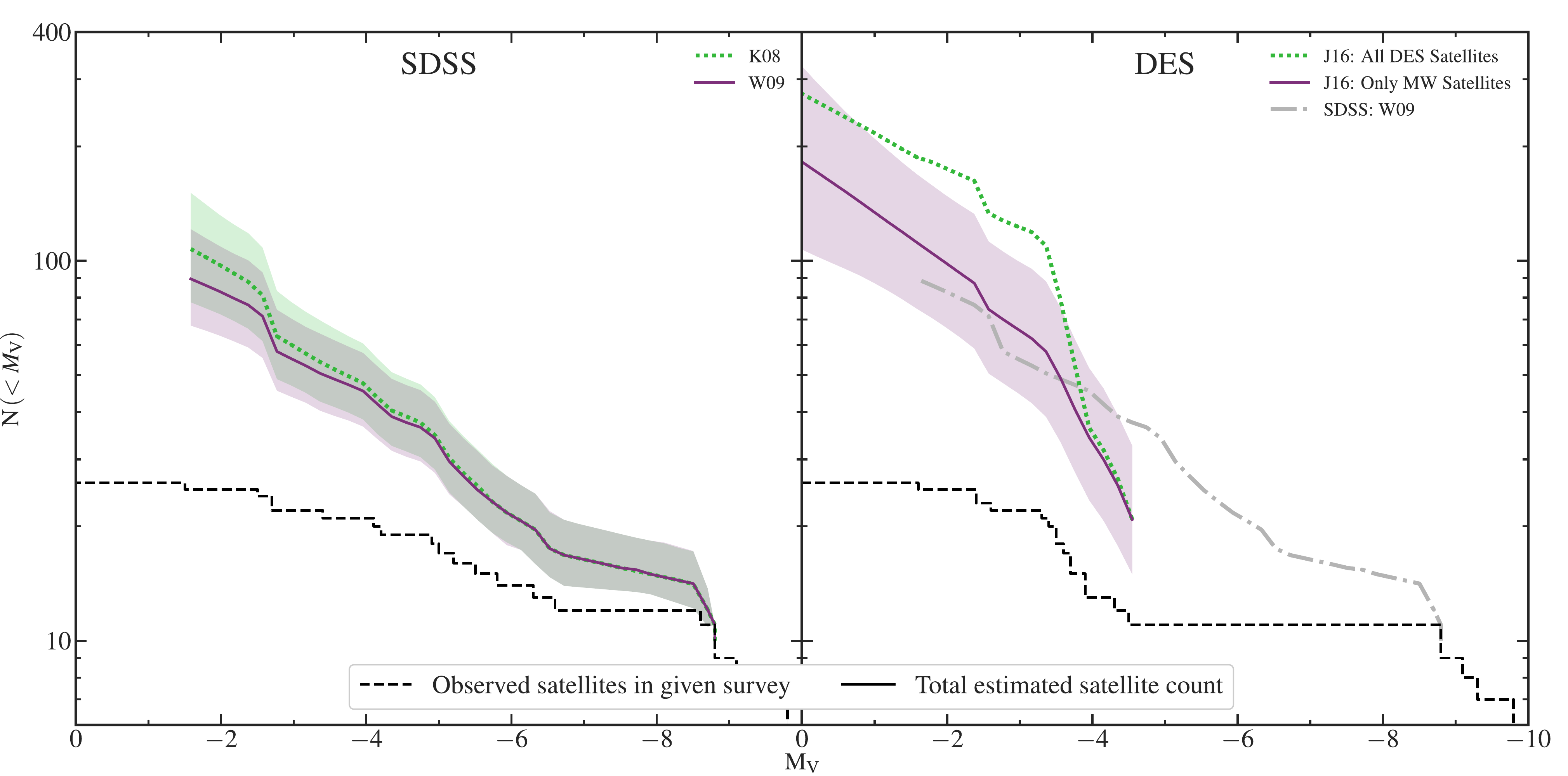}
	\vspace{-15pt}
	\caption{The total MW satellite galaxy luminosity functions
          inferred from the \SDSS{} and \DES{} surveys (left and right
          panels, respectively). The solid lines and corresponding
          shaded regions show the median estimates and associated
          $68$~per cent uncertainties. The dashed lines indicate the
          number of observed satellites within \Rout{} in each of the
          two surveys; these are input into the Bayesian inference
          method. For the \SDSS{}, we show estimates using the
          response functions of the two search algorithms devised by
          \citet[K08]{koposov_luminosity_2008} and
          \citet[W09]{walsh_invisibles_2009}. Both algorithms detect
          the same number of satellites, but the latter probes down to
          fainter magnitudes. For \DES{}, we use the
          \citet[J16]{jethwa_magellanic_2016} response function. This result
          is truncated at $\MV{}\leq-4.5$ as no satellites brighter than this
          have been observed in \DES{} within \kpc[300]. The \DES{}
          estimate (solid line) accounts for the possibility that some
          objects observed by \DES{} may be satellites of the
          \LMC{}. For reference, we also plot a second estimate which
          assumes that all \DES{} objects are associated with the MW
          (dotted line), as well as the \SDSS{}
          \citetalias{walsh_invisibles_2009} result (dot-dashed line).  }
	\label{fig:Individual_SDSS_DES_results}
	\vspace{-10pt}
\end{figure*}

There are two main factors that introduce uncertainties. First, the
distribution of satellites is not isotropic but flattened. As a result,
surveying different regions of the halo can introduce variations in the number
of observed objects. Secondly, the presence or absence of satellites in the
observed volume is a stochastic process. Given $N$~satellites and the
probability, $\eta$, of a satellite being inside the survey volume, then the
number of observed satellites in the survey is a binomial distribution with
parameters $N$ and $\eta$. To determine which of the two effects is dominant, we
applied the Bayesian inference method to the original subhalo distribution of
the Aq-A1 halo and to many isotropized versions of it. These were generated
keeping the same radial distances and isotropizing the angular coordinates. The
results of this test, presented in \figref{fig:SDSS_ido}, show that while
anisotropy makes a noticeable contribution to the uncertainty at faint
magnitudes, stochastic effects are the dominant source of uncertainty.

The \citetalias{tollerud_hundreds_2008} method accounts for anisotropy, but it
does not account for stochastic effects, which leads to an
underestimation of the errors. This underestimate is clearly seen in the mock
observation tests detailed in \appref{app:Testing_prev_methods}, where we find
that most of the \citetalias{tollerud_hundreds_2008} estimates lie further than
the $68$~per cent uncertainty interval from the input `true' luminosity function.
Given the probability, $\eta$, that a satellite is inside the volume
$V_{\rm eff}$, the \citetalias{tollerud_hundreds_2008} method predicts
$\eta^{-1}$ satellites within the halo---see \eqnref{eq:T08_estimate}
without the $\epsilon$ term. While this is true on average, for any
realization the number of satellites in the halo is given by a
negative-binomial distribution with mean value $\eta^{-1}$. The width of
this distribution, which characterizes the size of the stochastic effects,
gives rise to an additional uncertainty that is not included in the
\citetalias{tollerud_hundreds_2008} methodology.

\section{Results}
\label{sec:Results}
We now provide the results of our analysis using the \Aquarius{}
haloes rescaled to a fiducial MW halo mass of
$\Msun[1.0\times10^{12}]$ and within a fiducial radius,
$\router{}{=}\Rout{}$. Initially, we perform our analysis for the
\SDSS{} and \DES{} data separately, each requiring
extrapolations over large unobserved volumes.  Combining both surveys
reduces the uncertainty because of the larger volume coverage.  We
also address other issues, for example, the dependence of the inferred
total luminosity function on the assumed MW halo mass and on radial
distance.

\subsection{Separate estimates from SDSS and DES}
\label{sec:Separate_estimates}
The results of applying our Bayesian inference method to the
\SDSSDR{9} data set are displayed in the left-hand panel of
\figref{fig:Individual_SDSS_DES_results}. Also plotted here is the luminosity
function of all satellite galaxies observed in the \SDSSDR{9} survey
for which absolute magnitude measurements have been published to date;
these data are provided in \tabref{tab:obs_sat_table}. We adopt the
response functions of the two search algorithms detailed in
\secref{sec:Data}, by \citetalias{koposov_luminosity_2008} and
\citetalias{walsh_invisibles_2009}. The counts inferred using the \citetalias{koposov_luminosity_2008}
function are systematically higher than those obtained using the \citetalias{walsh_invisibles_2009}
function at absolute magnitudes fainter than $\MV{} \approx
-5.5$. This is expected and is a consequence of both algorithms
detecting the same number of satellites, but the
\citetalias{walsh_invisibles_2009} algorithm probing
deeper at fainter magnitudes. The larger scatter in the
\citetalias{koposov_luminosity_2008} estimate
reflects the additional uncertainty introduced by requiring an
extrapolation over larger volumes of the halo. In the remainder of
this paper we will use the results obtained using the
\citetalias{walsh_invisibles_2009} algorithm as
it is able to detect---at least in principle---fainter objects.

Down to magnitude $\MV{}{=}-2.7$ (corresponding to the faintest satellite
considered by \citeauthor{tollerud_hundreds_2008}), the \SDSS{} data imply
that there are at
least \medSDSStollimresult{}~(\CL{98\ {\rm per\ cent}}, statistical
error---note that the \CL{68\ {\rm per\ cent}} is
shown in the figure) dwarf galaxies within a radial distance of \Rout{}.
This is significantly lower than the estimate by
\citeauthor{tollerud_hundreds_2008}, who inferred \tollerudresult{} at
\CL{98\ {\rm per\ cent}}. The \citeauthor{tollerud_hundreds_2008}
estimate is higher for two reasons.
First, they adopted the \citetalias{koposov_luminosity_2008}
response function which is shallower than the
\citetalias{walsh_invisibles_2009} function. 
Secondly, their estimates were based on the \SDSSDR{5} data release
that observed $10$ satellites over a footprint of ${\sim}8000$ square
degrees. Since then, while \SDSSDR{9} has added an additional ${\sim}6500$
square degrees of sky coverage, it has detected only four new satellites
brighter than $\MV{}{=}-2.7$.

The result of applying our method to the \DES{} is shown in the
right-hand panel of \figref{fig:Individual_SDSS_DES_results}; in this case we
adopt the \citet{jethwa_magellanic_2016} response function. No
satellites are detected in \DES{} with magnitude in the range
$-8.9\lesssim\MV{}\lesssim-4.5$, so we interpolate between the values
calculated at each end of the range. Including all the \DES{}
satellites in the inference method returns twice as many satellites
with $\MV{}{\lesssim}-4$ than inferred from the \SDSS{} satellites
alone. This discrepancy is caused by the \DES{} footprint being adjacent to
the two Magellanic Clouds which, models suggest, are on their first infall
\citep{kallivayalil_third-epoch_2013,jethwa_magellanic_2016}. If that
were the case, then it is likely that the two Magellanic Clouds would
have contributed their own complement of satellite galaxies. These are
not distributed uniformly over the sky, but are still clustered around
the Magellanic Clouds \citep{sales_clues_2011}. As many as half of the
satellites detected by \DES{} could have come from the \LMC{}
\citep{sales_satellite_2007,jethwa_magellanic_2016}. Failing
to account for these localized associations would lead to an
overestimate of the total Galactic satellite population. We adopt the
probabilities of association of each of the \DES{} objects with the
\LMC{} inferred by \citet{jethwa_magellanic_2016} and include an
additional step in our analysis: for each mock survey pointing, we
generate a Monte Carlo realization in which the \DES{} satellites are
assigned either to the MW or to the \LMC{} according to these
probabilities. Only the \DES{} satellites assigned to the MW are then
included in the Bayesian inference.

The right-hand panel of \figref{fig:Individual_SDSS_DES_results} shows the
satellite luminosity function accounting for the
association of some \DES{} satellites to the \LMC{}. This estimate is in
good agreement with the estimate from the \SDSS{} for $\MV{}{\lesssim}-4$.
The discrepancy at brighter magnitudes is due to the lack of detection in
the \DES{} survey of any satellites brighter than $\MV{}{=}-4.5$ within a
distance of $\Rout$. While \DES{} is deeper than \SDSS{}, it covers a
smaller area on the sky and thus, for $\MV{}{\lesssim}-5$ and
$\MV{}{\gtrsim}-0.5$, \DES{} samples a smaller effective volume than
\SDSS{} (see \figref{fig:Reffs}). Nonetheless, the luminosity function
inferred from \DES{} is generally consistent with that inferred from
\SDSS{}, given the large uncertainties in both estimates.

\subsection{Combined estimate from SDSS+DES}
\label{sec:SDSSDES_estimates}
The best estimate of the total satellite luminosity function is
obtained by combining the \SDSS{} and \DES{}. We modify the
analysis described in \secref{sec:Apply_bayes_frame} by including a
second conical region oriented relative to the first one such that it
reproduces the approximate orientation of the real \SDSS{} and \DES{}.
The \SDSS{} vector is used to define the pointing
`direction' of this configuration; it uniformly samples the sky as
before. The second vector---corresponding to the \DES{}---is
fixed at an angle of $120\degree$ relative to the \SDSS{} vector but
is allowed to rotate around it. For each \SDSS{} pointing a
configuration is generated and a combined \SDSS{}+\DES{} luminosity
function is calculated. In practice, this analysis corresponds to that of
a survey of effective volume,
$V_{\rm eff,\, \SDSS{}}+V_{\rm eff,\, \DES{}}$, consisting of two disjoint
regions. The analysis otherwise proceeds as before.

\begin{figure}
	\includegraphics[width=\columnwidth]{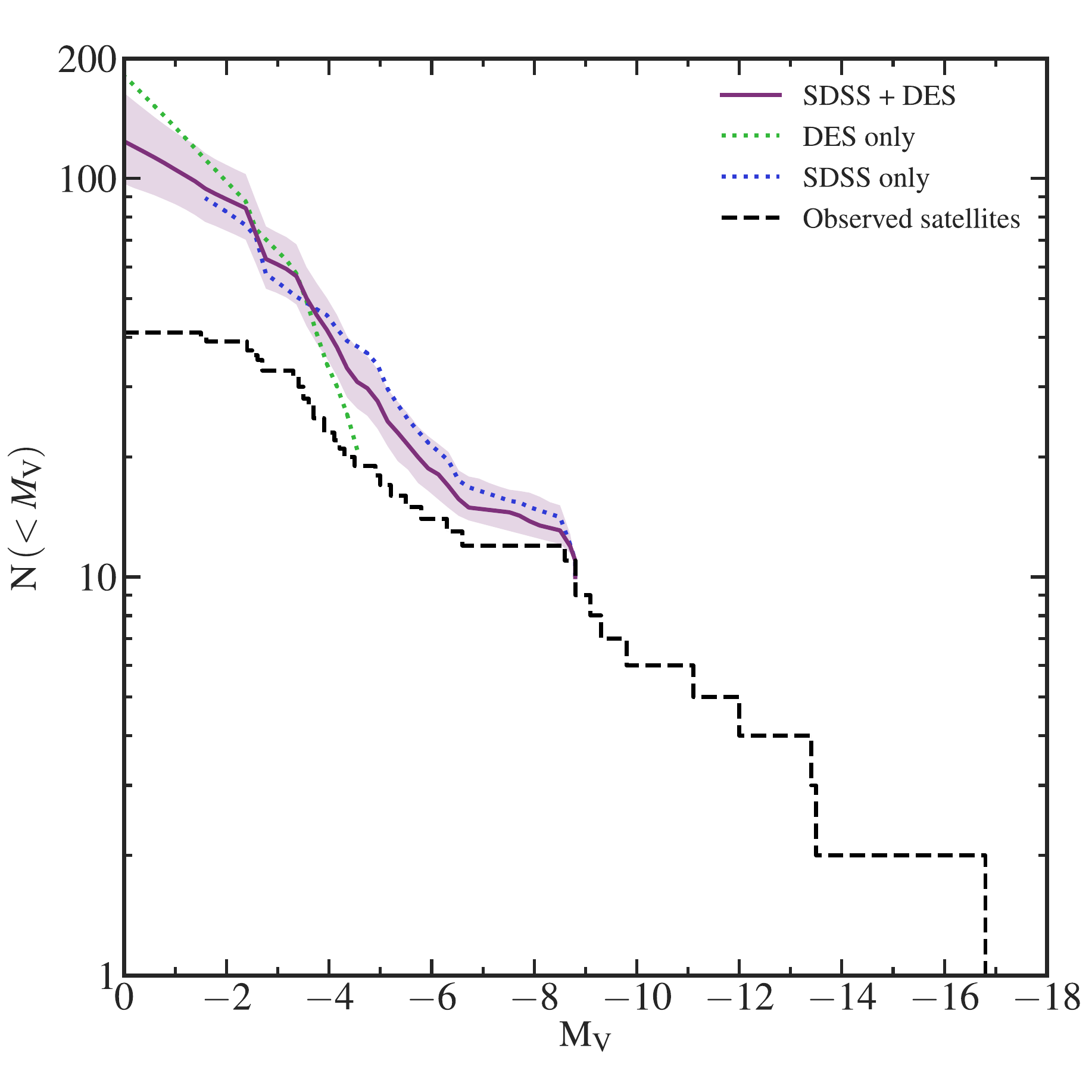}
	\vspace{-17pt}
	\caption{The total luminosity function of dwarf galaxies
          within a radius of \Rout{} from the Sun obtained
          from combining the \SDSS{} and \DES{} data. The
          solid line and the shaded region show the median estimate and
          its $68$~per cent uncertainty, respectively. The two dotted
          lines show the median satellite luminosity functions using
          \SDSS{} and \DES{} data separately. The luminosity function
          of all observed satellites within the \SDSS{} and \DES{}
          footprints inside \Rout{} is indicated by the dashed
          line. The total satellite luminosity function is well-fitted
          by the broken power law given in \eqnref{eq:Broken_power_law}.
        }
	\label{fig:Combined_SDSS_DES_result}
	\vspace{-10pt}
\end{figure}

The predicted total satellite luminosity function from the combined
\SDSS{}+\DES{} data is shown in \figref{fig:Combined_SDSS_DES_result}. This
estimate is consistent with those from the separate analyses of
\SDSS{} and \DES{} data: except in a few bins, the medians of the
individual estimates lie within the $68$~per cent uncertainty range of
the \SDSS{}+\DES{} estimate. When comparing with the combined result,
we find that the \SDSS{}-only estimate overpredicts the satellite
count for $\MV{}\le -4$, which is to be expected given that \DES{}
did not find any satellites brighter than $\MV{}{=}-4.5$ within our
fiducial radius of $\Rout$. In contrast, for $\MV{}>-4$, the
\SDSS{}-only estimate occasionally lies slightly below the total
satellite count, reflecting the large number of satellites with
$\MV{}\ge -4.5$ observed by \DES{}. The data associated with
\figref{fig:Combined_SDSS_DES_result} are provided in \tabref{tab:datatable} of
\appref{app:Data_table}.

We find that the total satellite luminosity function is well-fitted by the broken power law:
\begin{equation}
    \log_{10} N({<}\MV{}) =\begin{cases}  0.095\MV{}+1.85 & for\; \MV{}{<}-5.9 \\
                      0.156\MV{}+2.21 & for\; \MV{}{\geq}-5.9
       \end{cases}
    \label{eq:Broken_power_law} \;,
\end{equation}
that is, the faint end of the luminosity function is described by a
significantly steeper power law than the bright end.

\subsection{Dependence on the tracer population}
\label{sec:Tracer_dependence}
\begin{figure}
	\includegraphics[width=\columnwidth]{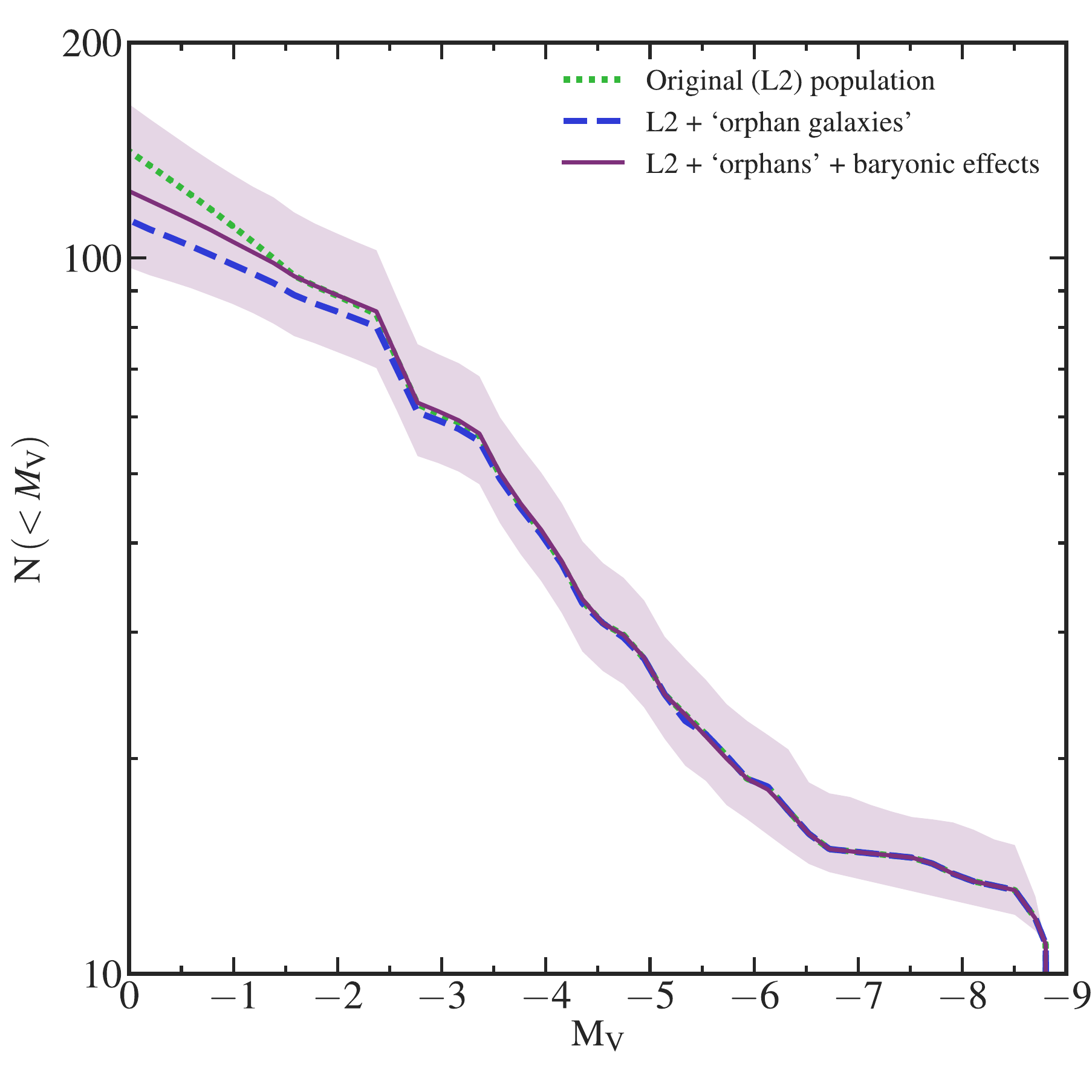}
	\vspace{-17pt}
	\caption{The sensitivity of the inferred satellite
            luminosity function to the two corrections applied to the
            subhalo population. The dotted line shows the inferred
            satellite count using the original subhalo distribution of
            \Aquarius{}. The dashed line shows the effect of adding
            subhaloes missing due to resolution effects, the so-called
            `orphan galaxies'. The solid line shows the results from 
            our analysis, in which we also account for subhalo
            depletion due to baryonic effects. The shaded region
            indicates the $68$~per cent uncertainty region of our final
            result.}
	\label{fig:compare_corrections}
	\vspace{-10pt}
\end{figure}

In \secref{sec:Tracer_population} we argued that in order to
make accurate predictions, it is necessary to incorporate two effects into
the analysis: the inclusion of unresolved subhaloes, i.e. `orphan
galaxies', and the depletion of subhaloes due to tidal disruption by the
central galaxy disc (i.e. baryonic effects). These changes primarily
involve the inner ${\sim}\kpc[50]$ of the  halo, the region to which the
faint end of the luminosity function is most sensitive.  Although
these two effects have opposite sign, they do not cancel
out completely. In \figref{fig:compare_corrections} we show the effect of each
of the two corrections, which are only important for the faintest
satellites $\left(\MV>-2\right)$.  Prior to any correction, the
$\MV{}{=}0$ satellite count is \origmedSDSSDESresult{}; the addition
of unresolved subhaloes reduces this to
\TtwomedSDSSDESresult{}. This is because the unresolved subhalo
population is very centrally concentrated; on average some
${\sim}85$~per cent of them lie within \kpc[50].  Accounting for subhalo
depletion due to baryonic effects produces a small upward shift in
the median to \medSDSSDESresult{}; a decrease of ${\sim}12$~per cent
relative to the uncorrected luminosity function inferred using the
L2 subhalo distribution of \Aquarius{} haloes.

\subsection{Dependence on the mass of the MW halo}
\label{sec:Halo_mass_influences}

\begin{figure}
	\includegraphics[width=\columnwidth]{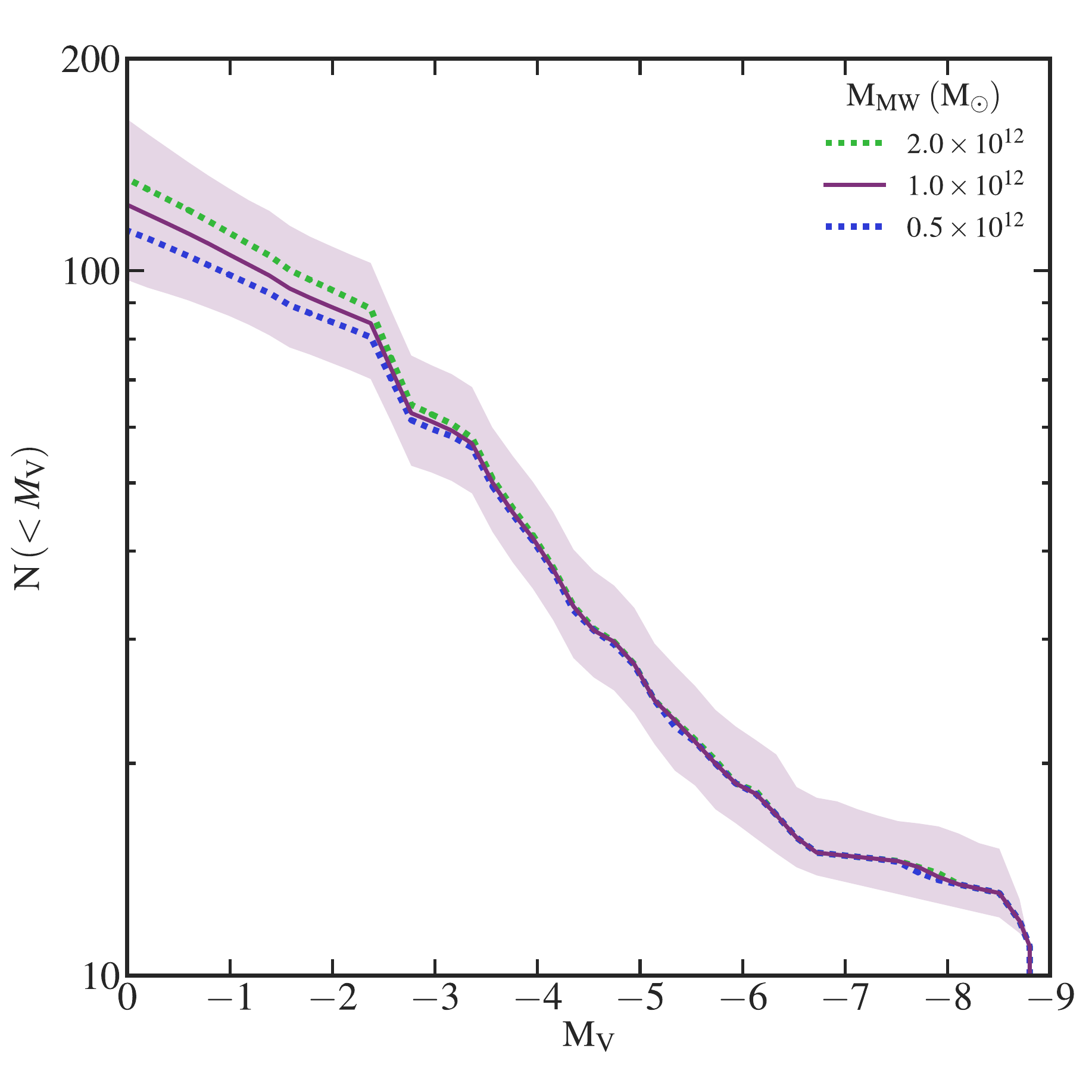}
	\vspace{-17pt}
	\caption{The dependence of the inferred total dwarf galaxy
          luminosity function within \Rout{} on the assumed mass of
          the MW halo. The lines show estimates for our fiducial MW
          halo mass of \Msun[1.0\times10^{12}] (used in previous
          plots) and for lighter and heavier MW haloes, as indicated
          in the legend. For the fiducial case, we show the
          median estimate (solid line) and the $68$~per cent uncertainty
          (shaded region). For the other two cases we show only the
          median estimates (dotted lines).  }
	\label{fig:halo_mass_variance}
	\vspace{-10pt}
\end{figure}

As we discussed in \secref{sec:mass_rescaling}, the MW halo mass is
poorly constrained, with recent estimates varying within a factor of
$2$ from our fiducial choice of $M_{\rm MW}=\Msun[1.0\times10^{12}]$
\citep[see the compilation of][]{wang_estimating_2015}. 
To investigate the sensitivity of the inferred total satellite
luminosity function to the MW halo mass, we repeated our analysis for two
extreme mass values, \Msun[0.5\times10^{12}] and
\Msun[2.0\times10^{12}], corresponding roughly to lower and upper
bounds for the MW halo mass \citep[e.g.][]{wang_estimating_2015}. To obtain
estimates for these halo masses, we rescaled the fiducial radial distribution of
subhaloes using the procedure described in \secref{sec:mass_rescaling}. The
inferred dwarf galaxy luminosity functions are displayed in
\figref{fig:halo_mass_variance}, which shows that despite the factor of 4
difference between the lowest and highest halo masses considered, no large
discrepancies begin to emerge until $\MV{}{\geq}-2.5$. Even at fainter
magnitudes, the differences are well within the $68$~per cent uncertainty
range for a given MW halo mass.

The number of subhaloes in a DM halo scales strongly with halo mass
\citep[e.g.][]{wang_missing_2012,cautun_subhalo_2014}, so naively we might
assume that the inferred satellite count follows the same relation. As
\figref{fig:halo_mass_variance} demonstrates, that is not the case; we see only
a weak variation of $N_{\rm tot}$ with $M_{\rm halo}$. The inferred satellite
count depends only on the shape of the normalized radial profile of subhaloes,
and not on the \textit{total} number of subhaloes. When expressed in terms of
$r\, /\, \Rvir{}$, i.e. radial distance in units of the virial radius of the halo, the
radial profile is largely independent of host mass
\citep{springel_aquarius_2008,han_unified_2016,hellwing_copernicus_2016}. Different
host masses correspond to different values of \Rvir{}, and thus any
features in the radial profile are mapped on to different physical
radial distances. If the radial distribution of subhaloes were a power
law, then the inferred satellite count would be independent of halo
mass: for fixed $r$, changing \Rvir{} would only lead to a shift in
the normalization of the radial profile, which is unimportant for our
analysis.

\subsection{Dependence on the outer radius cut-off}
\label{sec:Outer_radii_predictions}
\begin{figure}
	\includegraphics[width=\columnwidth]{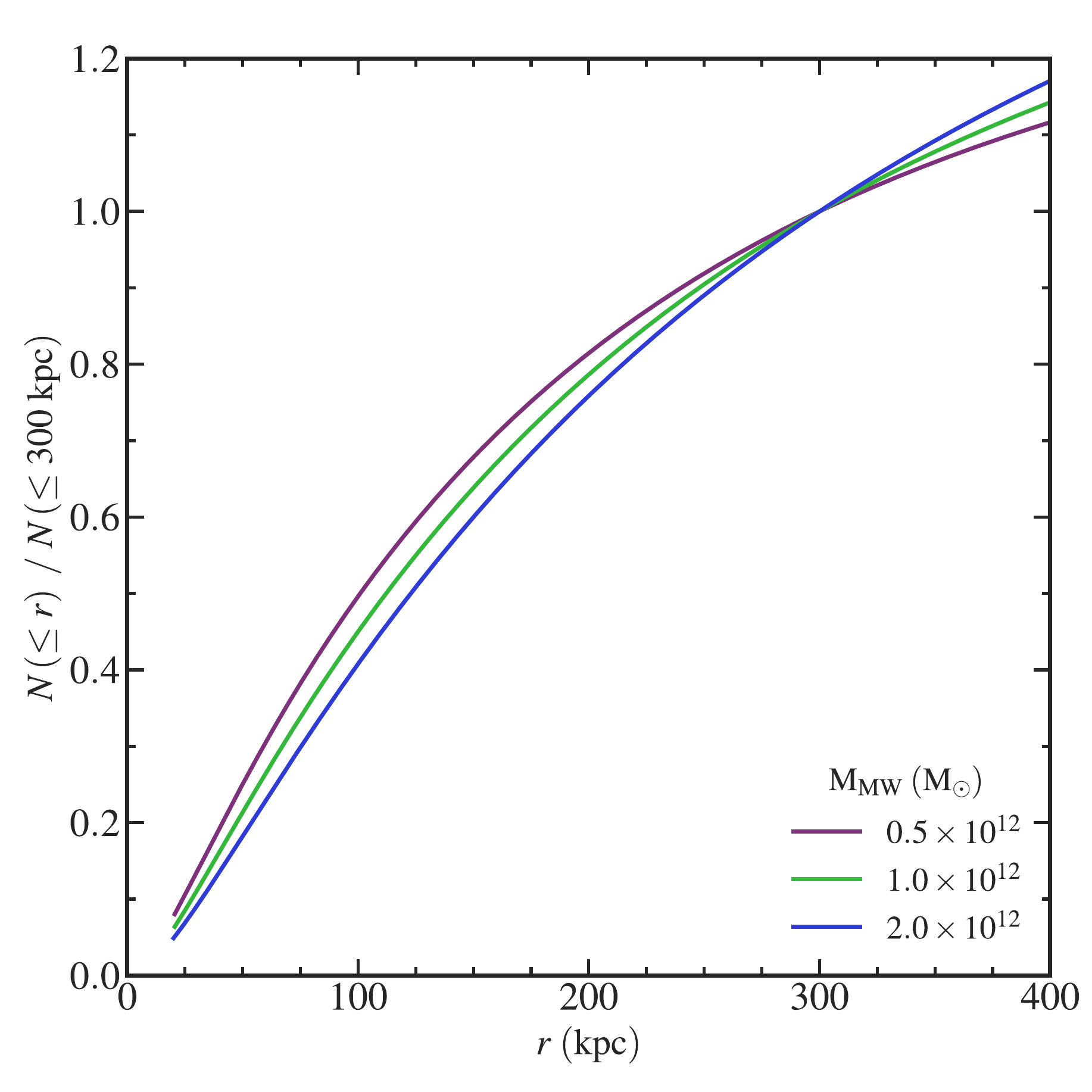}
	\vspace{-17pt}
	\caption{The radial dependence of the total number of
          satellites enclosed within radius $r$. The Y-axis gives the
          ratio of this number relative to the satellite count within
          \Rout{}, the fiducial radius used in this analysis. The
          result is independent of absolute magnitude, \MV{}, since
          subhaloes with different \vpeak{} cuts have the same radial
          profile. There is little dependence on the mass of the MW
          halo.  }
	\label{fig:n_sat_ratio}
	\vspace{-10pt}
\end{figure}

\figref{fig:n_sat_ratio} illustrates the dependence of the total
satellite count within a given radius, $r$, as a function of
$r$. These estimates follow from the observation that the radial
number density of subhaloes selected above a \vpeak{} threshold is
independent of the value of the threshold (see
\figref{fig:aq_ap-lum_comparison}), which suggests that the radial
distribution of satellites should also be independent of satellite
luminosity.

The fiducial radial distribution of subhaloes is well described by an Einasto
profile: the number of satellites within $\chi{=}r\, /\, \Rvir{}$ is given
by:
\begin{equation}
 	N\left(<\chi\right) = 4\pi\int_{0}^{\chi} \!n\left(\chi'\right) {\chi'}^2 \, \d{\chi'}
 \label{eq:number_satellites} \;,
\end{equation}
with $n\left(\chi'\right)$ the Einasto profile given by
\eqnref{eq:num_dens_prof_fit}. Performing the integration and
substituting for $\chi$ gives:
\begin{equation}
 	N\left(<r\right) = N\left(<\Rout{}\right)\; \frac{\LowIncGammaFunc{\dfrac{3}{\alpha}}{\dfrac{2}{\alpha}\left[\conc{}\chi\right]^\alpha}}{\LowIncGammaFunc{\dfrac{3}{\alpha}}{\dfrac{2}{\alpha}\left[\conc{}\dfrac{\kpc[300]}{\Rvir{}}\right]^\alpha}}
 	\label{eq:ratio_number_satellites} \;,
\end{equation}
where the function $\gamma$ is given by \eqnref{eq:Low_Inc_Gamma_Func}. The
radial dependence of $N\left(<r\right)$ is affected by the assumed
value for the MW halo mass through the dependence of \Rvir{} on halo
mass. \figref{fig:n_sat_ratio} shows the radial dependence of
$N\left(<r\right)$ for the three MW halo masses assumed in
\figref{fig:halo_mass_variance}; we find only a mild variation with MW
halo mass. Extending to distances farther than \Rout{} leads only to
modest increases in the satellite count, with an ${\sim}20$~per cent increase
at \kpc[400], which is roughly half way between the MW and M31. Of all
the satellites within \Rout{}, ${\sim}80$~per cent of them lie within
\kpc[200], the \Rvir{} value for a \Msun[1.0\times10^{12}] halo
mass. At even smaller radial distances, we find ${\sim}45$~per cent of the
satellites within \kpc[100].

\subsection{Apparent magnitude luminosity function}
\label{sec:App_mag_ests}
\begin{figure}
	\includegraphics[width=\columnwidth]{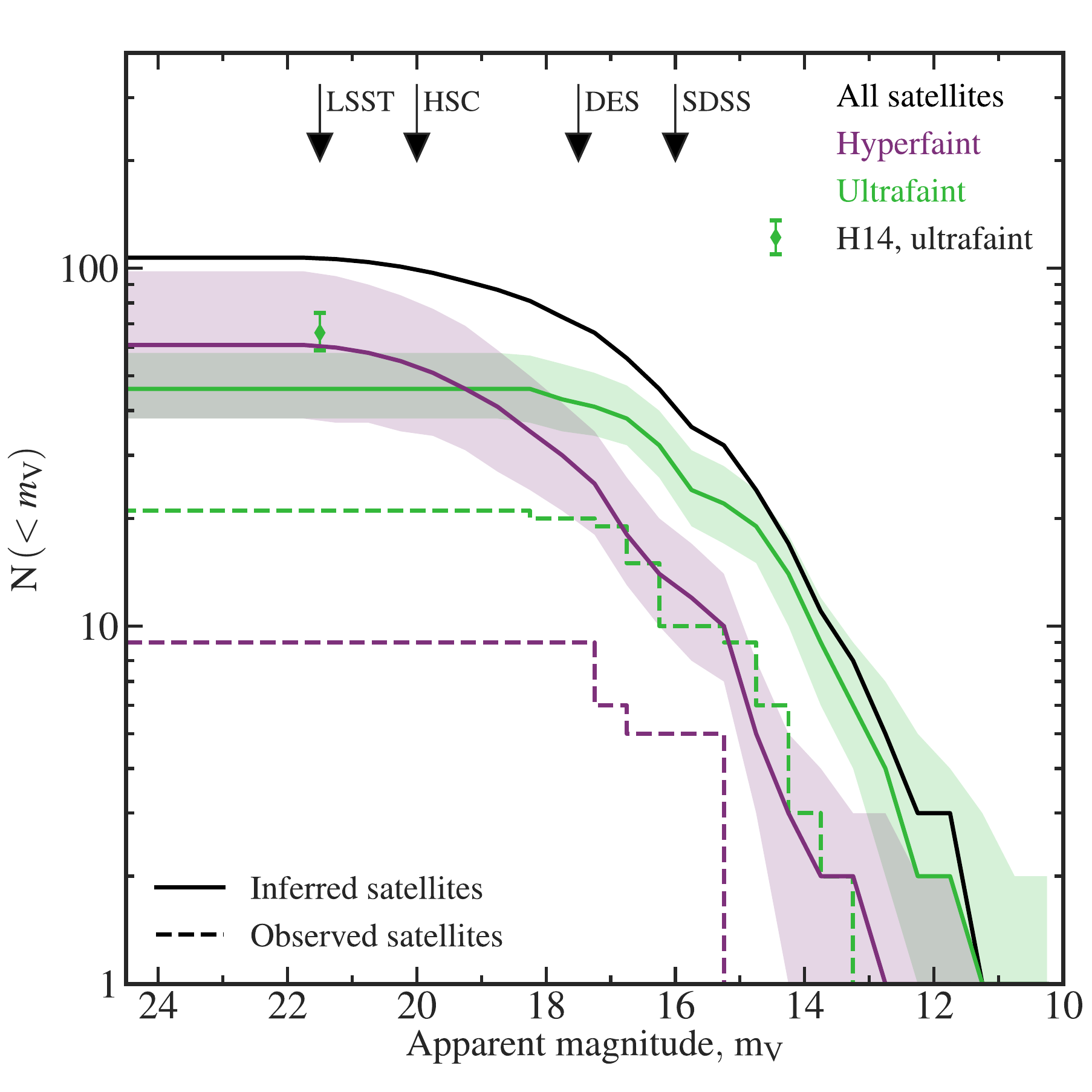}
	\vspace{-17pt}
	\caption{ The inferred Galactic satellite number counts within
          \kpc[300] as a function of apparent $V$-band magnitude,
          $m_{\rm V}$. The satellites are split into ultra- and
          hyperfaint dwarf galaxies, which correspond to objects with
          absolute magnitude in the range $-8 < \MV{} \leq -3$ and $-3
          < \MV{} \leq 0$, respectively. The solid lines display the
          median prediction, with the corresponding shaded regions
          indicating the $68$~per cent uncertainties. For reference the sum of the
          median predictions of both populations is also provided (black line).
          The diamond and associated
          error bars represent the \citet[H14]{hargis_too_2014} prediction and
          $68$~per cent uncertainty region for the total expected number of
          ultrafaint satellites. As before, the dashed lines display number
          counts of observed ultra- and hyperfaint dwarf galaxies within the
          \SDSS{} and \DES{}. The vertical arrows indicate the
          faintest satellites that can be detected in past and future
          surveys: \SDSS{} ($m_{\rm V}=16.0$), \DES{} ($m_{\rm V}=17.5$),
          \HSC{} ($m_{\rm V}=20.0$) and \LSST{} ($m_{\rm V}=21.5$).  }
	\label{fig:ult_vs_hyp}
	\vspace{-10pt}
\end{figure}

In this subsection we examine the prospects for discovery of faint
satellites in future surveys of the MW. For simplicity we
assume that the only factor that determines the detectability of a
satellite is its apparent luminosity, rather than its size or surface
brightness. We can then calculate the number counts of satellites as
a function of $V$-band magnitude.  To estimate apparent magnitudes, we
assign an absolute magnitude, $\MV{}$, to subhaloes by sampling the
inferred luminosity function from \secref{sec:SDSSDES_estimates},
i.e. the combined \SDSS{}+\DES{} estimate. We then use the subhalo
distance from the halo centre to compute the distance modulus and thus
the apparent magnitude. This process is repeated for the luminosity
functions generated from each pointing and observer location
combination---$6000$ in all. The results presented in this section
are for a MW halo mass of \Msun[{1.0\times10^{12}}] and for a \Rout{}
outer radius.

Dwarf galaxy counts as a function of apparent magnitude are shown in
\figref{fig:ult_vs_hyp}, where we split the population into two
classes: ultrafaint and hyperfaint dwarf galaxies, which we define as
objects in the absolute magnitude ranges: $-8 < \MV{} \leq -3$ and
$-3 < \MV{} \leq 0$ respectively. Within \Rout{} from the MW, we
expect to find \medufresult{}~(\CL{68\ {\rm per\ cent}}, statistical error) ultrafaint and
\medhfresult{}~(\CL{68\ {\rm per\ cent}}, statistical) hyperfaint dwarfs. The first number can
be compared to the slightly higher estimate of \hargisufresult{}~(\CL{68\ {\rm per\ cent}})
ultrafaints provided by \citet{hargis_too_2014}, based
solely on data from \SDSSDR{8}. We showed in
\figref{fig:Combined_SDSS_DES_result} that this population is usually
overestimated in predictions based only on \SDSS{} because of a higher
abundance of ultrafaint satellites in the \SDSS{} field than would be
expected from the total observed population. As discussed in
\secref{sec:compare_methods}, their uncertainties are also $28$~per cent too
small as stochastic effects were not accounted for in their
estimate. Most ultrafaints have apparent magnitudes brighter
than $18$, so surveys just $0.5$ magnitudes deeper than \DES{}---which
can detect satellites down to $m_V=17.5$---should be deep
enough to observe most ultrafaint dwarfs in the MW. The
luminosity function of hyperfaint dwarfs extends much fainter, with
most satellites having $m_V<21.5$. Discovering these would require a
survey \mgn{4} deeper than \DES{}; the Large Synoptic Survey
Telescope~(\LSST{}) is one such future survey. An all-sky \DES{}-like survey
would only lead to the detection of ${\sim}30$ hyperfaint dwarfs, a factor of
$4$ more than the currently known population.

\section{Discussion}
\label{sec:Discussion}
We have made new predictions for the total MW satellite
luminosity function by extrapolating the numbers of satellites
currently known using a new Bayesian inference method.  As input data
we use a combination of the recently discovered satellites
in the DES{} and the population previously known from
\SDSSDR{9}. As a prior for the radial distribution of the MW
satellites, which is needed for the extrapolation, we use the radial
distribution of subhaloes in the \Aquarius{} simulations of galactic
haloes having peak maximum circular velocity, \vpeak{}, above a given
threshold. We correct the subhalo distribution for unresolved subhaloes and account for subhalo depletion due to tidal disruption by the central disc. 
We showed in \figref{fig:radial_distribution_MW} that the radial distribution of
\vpeak{}-selected subhaloes provides a good match to that of the observed MW
satellites. We improve upon previous studies by introducing a new
Bayesian inference method, which overcomes the limitations of earlier
approaches. We also explore the effect of uncertainties in the MW halo
mass and derive a relation for rescaling our estimates to different
radii.

We find that, for a \Msun[1.0\times10^{12}] MW halo, there are
\medSDSSDESresult{}~(\CL{68\ {\rm per\ cent}}, statistical error) satellites brighter than
$\MV{}{=}0$ within \Rout{} of the Sun, which is slightly inconsistent with the
result from \citet{hargis_too_2014}. Our estimate is consistent with that of
\citet{jethwa_magellanic_2016} when adjusted for differing outer
radii; their estimate lies at the upper end of our
$68$~per cent uncertainty range. Our lower estimate is due to the inclusion of orphan
galaxies and baryonic effects, which decrease the inferred count of MW
satellites (see \figref{fig:compare_corrections}).
Compared with the \citet{tollerud_hundreds_2008}
estimate of \tollerudresult{}~(\CL{98\ {\rm per\ cent}}) satellites brighter than
$\MV{}{=}-2.7$ within \Rout{}, our estimate of
\medSDSSDEStollimresult{}~(\CL{98\ {\rm per\ cent}}, statistical) is a factor of
${\sim}\tollresultfrac{}$ lower. The origin of this discrepancy is
primarily the use by \citeauthor{tollerud_hundreds_2008} of the shallower
\citetalias{koposov_luminosity_2008} response function as opposed to the
\citetalias{walsh_invisibles_2009} function that we use here. Furthermore,
since their work the \SDSS{} survey footprint has increased in size by
${\sim}80$~per cent, while the number of discovered satellites inside this footprint
has increased by very little.
We also note that previous studies have underestimated their
uncertainty ranges because they have not properly accounted for
stochastic effects, which are broadly independent of satellite brightness (see
\secref{sec:compare_methods} for a more in-depth discussion).

The future detection of dwarfs depends on their apparent magnitude and
we can estimate the luminosity thresholds that future surveys will
need to exceed in order to detect the satellite population inferred
in this study. In our total inferred population there are
\medufresult{}~(\CL{68\ {\rm per\ cent}}, statistical) ultrafaint dwarf galaxies (with
magnitudes in the range $-8 < \MV{} \leq -3$), of which ${\sim}20$
have been observed so far. We find that the majority of these have
apparent magnitudes brighter than $\mV{}{=}18$; these would be
discoverable with surveys just $0.5$ magnitudes deeper than
\DES{}. There are ${\sim}30$ such dwarfs still to be discovered in the
MW, of which ${\sim}7$ should lie inside the \SDSSDR{9} footprint but
beyond its detection limit.  Our \medhfresult{}~(\CL{68\ {\rm per\ cent}}, statistical)
hyperfaint dwarfs (with magnitudes $\MV\geq-3$) make up some $62$~per cent of
our total population and have apparent magnitudes brighter than
$\mV{}{=}21$; discovering these would require a survey \mgn{4}
deeper than \DES{}. The planned \LSST{} survey should cover
approximately half of the sky and will therefore be able to find
half of the inferred count of \medhfresult{} hyperfaint dwarfs. The
sizes of both populations are slightly inconsistent with the lower end of
estimates by \citet{hargis_too_2014}.

Our inferred satellite galaxy luminosity function likely represents a
lower limit to the true population. Our method takes the observed
satellites, which are found in surveys with various detectability
limits, as a sample of the global population. In particular, the
observed surface brightness cut-off suggests that there could be a
population of faint, spatially extended dwarfs that are inaccessible
to current surveys \citep[e.g. see][]{torrealba_feeble_2016}. To account
for this in our method would require deeper observations than are currently
available.

A further complication arises from the presence of the \LMC{}, which,
given its large mass, is likely to have brought its own complement of
satellites. The \LMC{} may be on its first infall
\citep{sales_clues_2011,kallivayalil_third-epoch_2013,
  jethwa_magellanic_2016} and the spatial distribution of the
satellites it brought with it could be very anisotropic
\citep{jethwa_magellanic_2016}. While we accounted for the probability
that a large fraction of \DES{} detections may be associated with the
\LMC{}, our analysis does not account for the presence of \LMC{}
satellites outside the \DES{} footprint. To do so would require a
prior on the present-day spatial distribution of \LMC{}
satellites. Before infall, the \LMC{} could have had perhaps as much
as a third of the MW satellite count \citep{jethwa_magellanic_2016},
though this estimate is very uncertain due to poor constraints on the
MW and especially the \LMC{} halo mass. At face value, this
could add at most ${\sim}50$ satellites to the total count.

Inherent to all analyses that estimate the satellite luminosity function are
several systematics which, with a few exceptions, mainly affect the faint end of
the luminosity function. The most important of these is the assumed radial
distribution of subhaloes, which needs to be determined from
cosmological simulations. We showed that the distribution of \vpeak{}-selected
subhaloes matches both the luminosity-independent radial distribution of
observed MW satellites and that of
state-of-the-art hydrodynamic simulations such as \Apostle{} (see
\figrefs{fig:aq_ap-lum_comparison}{fig:radial_distribution_MW});
consequently, we think that any systematic effect on the inferred satellite
count arising from our choice of fiducial tracer population is likely to be
small. To obtain our fiducial subhalo sample, we needed to correct for two
effects that are not well understood. Even the highest resolution simulations,
such as those of the \Aquarius{} project, can suffer from resolution
effects, particularly near the centre of the host halo. This issue is
common to \textit{all} cosmological simulations, and we addressed it by
including `orphan galaxies' (i.e. galaxies whose haloes have been disrupted)
identified by applying the Durham semi-analytic model of galaxy formation,
\Galform{}, to the \Aquarius{} simulations. This effect is only significant for
the faint end of the satellite luminosity function ($\MV{} \gtrsim -3$) since
${\sim}85$~per cent of the orphan population lies within \kpc[50] of the centre, the
region to which the faint end is most sensitive. We also accounted for
baryonic effects on the subhalo mass function by lowering its
amplitude in accordance with the prescription in
\appref{app:Baryonic_effects}, using depletion factors based on the \Apostle{} project
\citep{sawala_shaken_2017}. \citet{garrison-kimmel_not_2017} argued for a larger depletion in the inner ${\sim}\kpc[30]$ than \citeauthor{sawala_shaken_2017}, while \citet{errani_effect_2017} claim that, due to their limited resolution, most simulations overpredict the subhalo depletion factor. As discussed in
\secref{sec:Tracer_dependence}, although this correction introduces
noticeable changes in the predicted satellite luminosity function,
these lie within our error bounds, and are smaller in magnitude than
those introduced by the addition of orphan galaxies. These changes
primarily affect the faint end of the satellite luminosity function
above $\MV{\geq}-2$, which is also the most theoretically and
observationally uncertain part of the luminosity function
independently of these effects.

A second important systematic is the choice of observed satellite
population. In this work we used satellites discovered in the \SDSS{} and
\DES{}. Although all satellites in the former have been
spectroscopically confirmed as DM-dominated dwarf galaxies, over three-quarters
of the \DES{} satellites have not (yet). We choose to use
\textit{all} \DES{} satellites in our analysis. This is motivated by
considering the size-magnitude plane
\citep[e.g.][fig.~$4$]{drlica-wagner_eight_2015} that shows that most \DES{}
satellites are more consistent with the properties of Local Group galaxies
than with the population of known globular clusters. Reclassifying some of the
\DES{} detections as globular clusters would lower the inferred total satellite
count at the faint end of the luminosity function ($\MV{} \geq -4$), but would
not affect the bright end. Given the good agreement between the \SDSS{}-only
and \DES{}-only estimates of the total satellite count, we predict that most
\DES{} detections are dwarf galaxies.

The mass of the MW halo is poorly constrained. However,
the inferred satellite luminosity function is largely independent of the host
halo mass, except at magnitudes fainter than $\MV{}{=}-3$ where it shows a very
weak mass dependence (see \figref{fig:halo_mass_variance}). Instead of
marginalizing over the MW halo mass distribution, we provide a
means of converting between halo masses at the extremes of the range of constraints.

The MW is the smaller partner of a paired system, which could introduce
anisotropies into the MW's substructure due to interactions with M31; these
would be manifest in the form of more correlated structure. Our choice of
\Rout{} for our fiducial radius is less than the midpoint of the MW-M31
distance, minimizing any effects from interactions with M31 and allowing us to
model the MW approximately as an isolated halo. In addition, this value is often
used in the literature \citep[e.g.][]{hargis_too_2014,jethwa_magellanic_2016}
and is close to the expected virial radius of the MW halo. Our choice of
fiducial radius should not be interpreted as precluding the eventual discovery
of other satellites further out than this.

The dependence of the total satellite count on MW halo mass is not determined by
the number of subhaloes at fixed mass, but by the shape of the normalized
subhalo radial number density profile. A weak halo mass dependence
arises from the non-power law nature of the subhalo radial profile:
features in this profile are remapped to different physical distances
for different halo masses, resulting in a variation in the predicted
luminosity function. As a direct consequence, this implies that
changes in the assumed MW halo mass, which determines the number of DM
substructures, alter the abundance matching relation for
Galactic dwarfs; in this regime not all subhaloes of a given mass host a
visible galaxy \citep{sawala_bent_2015}. We find that doubling the halo
mass roughly doubles the number of subhaloes
\citep{wang_missing_2012,cautun_subhalo_2014}, so that there are more
of them at fixed \vpeak{}. A more massive MW halo would then require the
same dwarfs to be placed in subhaloes with higher \vpeak{} than they
would for a lower MW mass halo. 

The spatial distribution of subhaloes---upon which our predictions
rely---is partly determined by cosmology but is also affected by the
internal dynamics of haloes. In turn, these are influenced by the mass
function of subhaloes and their accretion rate, both of which are
fairly universal in both \LCDM{} and WDM models
\citep{springel_aquarius_2008,ludlow_massconcentrationredshift_2016}. Recent work by
\citet{bose_substructure_2017} has shown that the radial distribution
of subhaloes is broadly independent of the nature of the DM.
Our predictions are therefore applicable to other DM models and can,
in fact, be used to constrain the masses of WDM particles.

\section{Conclusions}
\label{sec:Conclusions}
An estimate of the MW's complement of satellite galaxies is
required until deeper, more complete surveys that could
discover more faint galaxies are undertaken in the next few years.
These predictions can be used to address numerous
outstanding astrophysical questions, from understanding the effects of
reionization on low mass haloes, to constraining the properties of dark matter
particles.

In this work we have, for the first time, combined data from \SDSS{}
and \DES{}---which together cover nearly half of the sky---to infer
the MW's full complement of satellite galaxies. Our method requires a
prior for the radial distribution of satellites, which we obtain from
the subhalo populations of the \Aquarius{} suite of high-resolution
DM-only simulations in which we account for the competing effects of
resolution and subhalo depletion due to interaction with the central baryonic
disc (see \secref{sec:Discussion}). We have shown that selecting subhaloes by
their peak maximum circular velocity provides a good match to the
radial distribution of observed MW satellites
(see~\figref{fig:radial_distribution_MW}).

The Bayesian method we have introduced to make these estimates
overcomes some of the limitations of previous analyses (see
\figref{fig:aq_extrap_methods}), and properly accounts for
stochastic effects. For each observed dwarf galaxy, the method
estimates how many objects are needed to find one such satellite in
the survey volume. These results are averaged over multiple DM haloes
to characterize uncertainties arising from halo-to-halo
variation.

Within \kpc[300] of the Sun---and assuming a MW halo mass of
\Msun[{1.0\times10^{12}}]---we predict that the MW has
\medSDSSDESresult{}~(\CL{68\ {\rm per\ cent}}, statistical error) satellites brighter than
$\MV{}{=}0$ (see \figref{fig:Combined_SDSS_DES_result}). Of these, we
expect to find \medufresult{}~(\CL{68\ {\rm per\ cent}}, statistical) ultrafaint dwarf galaxies
$\left(-8<\MV{}\leq-3\right)$, a result that is marginally inconsistent with the
lower end of the \citet{hargis_too_2014} estimate, but nearly a
factor of $5$ smaller than the \citet{tollerud_hundreds_2008}
estimate. All the Galactic ultrafaints could be detected by a survey
just $0.5$~magnitudes deeper than \DES{}. We also expect to find a
population of \medhfresult{}~(\CL{68\ {\rm per\ cent}}, statistical) hyperfaint dwarfs
$\left(-3<\MV{}\leq0\right)$, and to obtain a full census of this
population would need a survey \mgn{4} deeper than \DES{}. The
\LSST{} survey should be able to see at least half of this faint
population of dwarf galaxies in the next decade.

In all methods seeking to estimate the total luminosity function certain
assumptions must be made. In particular, an important assumption is the radial
distribution of the true satellite population, which is best inferred from a
cosmological simulation. Here, we have used a set of the highest resolution
DM-only simulations available and, most importantly, a method for selecting the
subhaloes that are expected to host satellites that has been shown to give
consistent results for a number of observed properties of the MW satellite
population, such as the radial distribution of and counts of bright observed MW
satellites. This does not guarantee that the extrapolation is free of
systematic effects but as \figref{fig:radial_distribution_MW} shows, in the
regime where we can check with available data, any such systematics are small.

The estimates above represent only lower limits to the total
number of Galactic satellites (see \secref{sec:Discussion}) because
they do not take into account very low surface brightness objects that
may have been missed in current observations. In addition,
the estimate does not account for some of the satellites brought in
by the \LMC{} which today lie outside the \DES{} footprint (which at
most would increase the total count by $30$~per cent).

While our key results assume a MW halo mass of
\Msun[{1.0\times10^{12}}], our analysis shows that the predicted dwarf
galaxy luminosity function is independent of host halo mass for
objects brighter than $\MV{}{=}-3$ (see
\figref{fig:halo_mass_variance}). For fainter satellites we find a
weak dependence on halo mass, with a more massive MW halo playing host
to more satellites. Our tests assuming extreme MW halo mass values
(\Msun[{[0.5,2.0]\times10^{12}}]) reveal that the resulting luminosity
functions lie well within the $68$~per cent uncertainty range calculated
for our fiducial MW halo mass. Of the dwarfs within our fiducial
distance of \Rout{}, ${\sim}45$~per cent and ${\sim}80$~per cent are found
within $100$ and \kpc[200], respectively.

The results of this study provide a useful reference point for
comparing theoretical predictions with the measured abundance of
satellite galaxies in the MW. However, it must be borne in mind that
the MW is only one system and that the abundance of satellites around
similar galaxies exhibits considerable scatter
\citep{guo_satellite_2012,wang_satellite_2012}.

The code that implements our method to estimate the total population of MW
satellite galaxies is available online \citep{newton_mw_2018}. In
addition, we also make available all data that are required to
reproduce our results (e.g.~\figref{fig:Combined_SDSS_DES_result}).

\section*{Acknowledgements}
The authors would like to thank the anonymous referee for detailed, insightful,
and thorough feedback that improved the quality of the manuscript. We would
also like to thank Till Sawala for useful discussions and for providing the raw
data used in \appref{app:Baryonic_effects}, and Roan Haggar for code-testing
the public software. This research made use of \numpy{} \citep{walt_numpy_2011},
\scipy{} \citep{jones_scipy_2011} and \matplotlib{}
\citep{hunter_matplotlib_2007}. ON was supported by the Science and Technology
Facilities Council~(STFC) through grant ST/N50404X/1 and MC, ARJ, and CSF were
supported by STFC grant ST/L00075X/1. This work used the DiRAC Data Centric
system at Durham University, operated by the Institute for Computational
Cosmology on behalf of the STFC DiRAC HPC Facility (\url{www.dirac.ac.uk}).
This equipment was funded by BIS National E-infrastructure capital grant
ST/K00042X/1, STFC capital grants ST/H008539/1 and ST/K00087X/1, STFC DiRAC
Operations grant ST/K003267/1, and Durham University. DiRAC is part of the
National E-Infrastructure.

%%%%%%%%%%%%%%%%%%%%%%%%%%%%%%%%%%%%%%%%%%%%%%%%%%



\bibliographystyle{mnras}
\bibliography{Milky_Way_Satellites}

%\begin{thebibliography}{99}
%\end{thebibliography}

%%%%%%%%%%%%%%%%%%%%%%%%%%%%%%%%%%%%%%%%%%%%%%%%%%

%%%%%%%%%%%%%%%%% APPENDICES %%%%%%%%%%%%%%%%%%%%%

\appendix
\newcommand{\satellitetablenotes}{
	\begin{tabular}{l}
	$^a$ Obtained from \citet[Fig.~9]{jethwa_magellanic_2016}.\\
	\multirow{2}{\columnwidth}{$^b$ The method of detection was different to that applied to other satellites in the \SDSS{} survey.}\\
	\\
	$^c$ Not spectroscopically confirmed.\\
	$^d$ No probability of association with \LMC{} provided.\\
	\multirow{3}{\columnwidth}{$^e$ Data reproduced from \citet[tables~2 and~3]{mcconnachie_observed_2012} unless indicated otherwise: (1)~\citet{kim_heros_2015, kim_portrait_2016}, (2)~\citet{watkins_substructure_2009}, (3)~\citet[Table~4]{drlica-wagner_eight_2015}, (4)~\citet{carlin_deep_2017}, (5)~\citet{li_farthest_2017}, (6)~\citet{koposov_kinematics_2015}, (7)~\citet{walker_magellan/m2fs_2016}.}\\
	\\
	\\
	\end{tabular}
}
\newcommand{\secondtablenotes}{
	\begin{tabular}{l}
	$^a$ Not spectroscopically confirmed.\\
	\multirow{5}{\columnwidth}{$^b$ Data reproduced from:
    (1)~\citet{torrealba_at_2016},
    (2)~\citet{torrealba_feeble_2016},
    (3)~\citet{laevens_sagittarius_2015},\\
    (4)~\citet{carlin_deep_2017},
    (5)~\citet{martin_hydra_2015},\\
    (6)~\citet{homma_new_2016}, (7)~\citet{homma_searches_2018},\\
    (8)~\cite{torrealba_discovery_2018},
    (9)~\citet{drlica-wagner_ultra-faint_2016}.}\\
	\\
    \\
	\\
	\end{tabular}
}
%\onecolumn
\section{Tables of known satellite galaxies}
\label{app:Known_satellites}
\FloatBarrier
\begin{table}
	\centering
	\caption{Known MW satellite galaxies identified in surveys used in this analysis, grouped according to the survey in which they were detected. For each satellite we provide its absolute $V-$band magnitude, \MV{}, heliocentric distance, ${\rm D_\odot}$, and -- for \DES{} satellites -- its probability of association with the \LMC{}.}
	\label{tab:obs_sat_table}
	\begin{tabular}{@{}lrrrl@{}} % four columns, alignment for each
		\hline
		\hline
		\multicolumn{1}{c}{Satellite} & ${\rm M_V}$ & \pbox{5cm}{\centering ${\rm D_\odot}$ (\kpc{})} & ${\rm p}_{\LMC{}}^a$ & Reference$^e$\\
		\hline
		\hline
		\multicolumn{5}{c}{Classical}                          \\
		Carina            &  -9.1 & 105 &             &        \\
		Draco I           &  -8.8 &  76 &             &        \\
		Fornax            & -13.4 & 147 &             &        \\
		Leo I             & -12.0 & 254 &             &        \\
		Leo II            &  -9.8 & 233 &             &        \\
		LMC               & -18.1 &  51 &             &        \\
		Ursa Minor        &  -8.8 &  76 &             &        \\
		SMC               & -16.8 &  64 &             &        \\
		Sculptor          & -11.1 &  86 &             &        \\
		Sextans           &  -9.3 &  86 &             &        \\
		Sagittarius I     & -13.5 &  26 &             &        \\
		\hline
		\multicolumn{5}{c}{\SDSSDR{9}}                         \\
		Bo\"{o}tes I      &  -6.3 &  66 &             &        \\
		Bo\"{o}tes II     &  -2.7 &  42 &             &        \\
		Canes Venatici I  &  -8.6 & 218 &             &        \\
		Canes Venatici II &  -4.9 & 160 &             &        \\
		Coma              &  -4.1 &  44 &             &        \\
		Hercules          &  -6.6 & 132 &             &        \\
		Leo IV            &  -5.8 & 154 &             &        \\
		Leo V             &  -5.2 & 178 &             &        \\
		Leo T             &  -8.0 & 417 &             &        \\
		Pegasus III       &  -3.4 & 215 &             & (1)    \\
		Pisces I$^b$      &   ... &  80 &             & (2)    \\
		Pisces II         &  -5.0 & 182 &             &        \\
		Segue I           &  -1.5 &  23 &             &        \\
		Segue II          &  -2.5 &  35 &             &        \\
		Ursa Major I      &  -5.5 &  97 &             &        \\
		Ursa Major II     &  -4.2 &  32 &             &        \\
		Willman I         &  -2.7 &  38 &             &        \\
		\hline
		\multicolumn{5}{c}{\DES{}}                             \\
		Cetus II$^c$         &   0.0 &  30 & $0.00^d$ & (3)    \\
		Columba I            &  -4.2 & 183 & 0.11     & (4)    \\
		Eridanus II          &  -7.1 & 366 & $0.00^d$ & (5)    \\
		Eridanus III$^c$     &  -2.4 &  95 & $0.00^d$ & (3)    \\
		Grus I$^c$           &  -3.4 & 120 & 0.64     & (3)    \\
		Grus II$^c$          &  -3.9 &  53 & 0.57     & (3)    \\
		Horologium I         &  -3.5 &  87 & 0.79     & (3,6) \\
		Horologium II$^c$    &  -2.6 &  78 & 0.80     & (3)    \\
		Indus II$^c$         &  -4.3 & 214 & 0.19     & (3)    \\
		Phoenix II$^c$       &  -3.7 &  95 & 0.75     & (3)    \\
		Pictoris$^c$         &  -3.7 & 126 & 0.62     & (3)    \\
		Reticulum II         &  -3.6 &  32 & 0.75     & (3,6) \\
		Reticulum III$^c$    &  -3.3 &  92 & 0.58     & (3)    \\
		Tucana II            &  -3.9 &  58 & 0.75     & (3,7) \\
		Tucana III$^c$       &  -2.4 &  25 & 0.52     & (3)    \\
		Tucana IV$^c$        &  -3.5 &  48 & 0.79     & (3)    \\
		Tucana V$^c$         &  -1.6 &  55 & 0.81     & (3)    \\
		\hline
		\hline
	\end{tabular}
	\satellitetablenotes{}
\end{table}

\begin{table}
	\centering
	\caption{Known MW satellite galaxies identified in surveys \textit{not} used in this analysis, grouped according to the survey in which they were detected. We provide the same data for each satellite as described in \tabref{tab:obs_sat_table}.}
	\label{tab:sec_sat_table}
	\begin{tabular}{lrrrl} % four columns, alignment for each
		\hline
		\hline
		\multicolumn{1}{c}{Satellite} & ${\rm M_V}$ & \pbox{5cm}{\centering ${\rm D_\odot}$ (\kpc{})} & Reference$^b$\\
		\hline
		\hline
		\multicolumn{4}{c}{\ATLAS{}}     \\
		Aquarius II          &  -4.2 & 108 &    (1) \\
		Crater II            &  -8.2 & 118 &    (2) \\
		\hline
		\multicolumn{4}{c}{\PAN{}}       \\
		Draco II             &  -2.9 &  20 &    (3) \\
		Sagittarius II$^a$   &  -5.2 &  67 &    (3) \\
		Triangulum II        &  -1.2 &  28 &    (4) \\
		\hline
		\multicolumn{4}{c}{\SMASH{}}     \\
		Hydra II             &  -4.8 & 134 &    (5) \\
		\hline
		\multicolumn{4}{c}{\HSC{}}       \\
		Virgo I$^a$          &  -0.3 & 91  &    (6) \\
		Cetus III$^a$        &  -2.4 & 251 &    (7) \\
		\hline
		\multicolumn{4}{c}{\Mag{}}       \\
		Carina II            &  -4.5 & 37  &    (8) \\
		Carina III$^a$       &  -2.4 & 28  &    (8) \\
		Pictoris II$^a$      &  -3.2 & 45  &    (9) \\
		\hline
		\hline
	\end{tabular}
	\secondtablenotes{}
\end{table}
\FloatBarrier
\section{Effects of resolution}
\label{app:Resolution_effects}

\begin{figure}
	\includegraphics[width=\columnwidth]{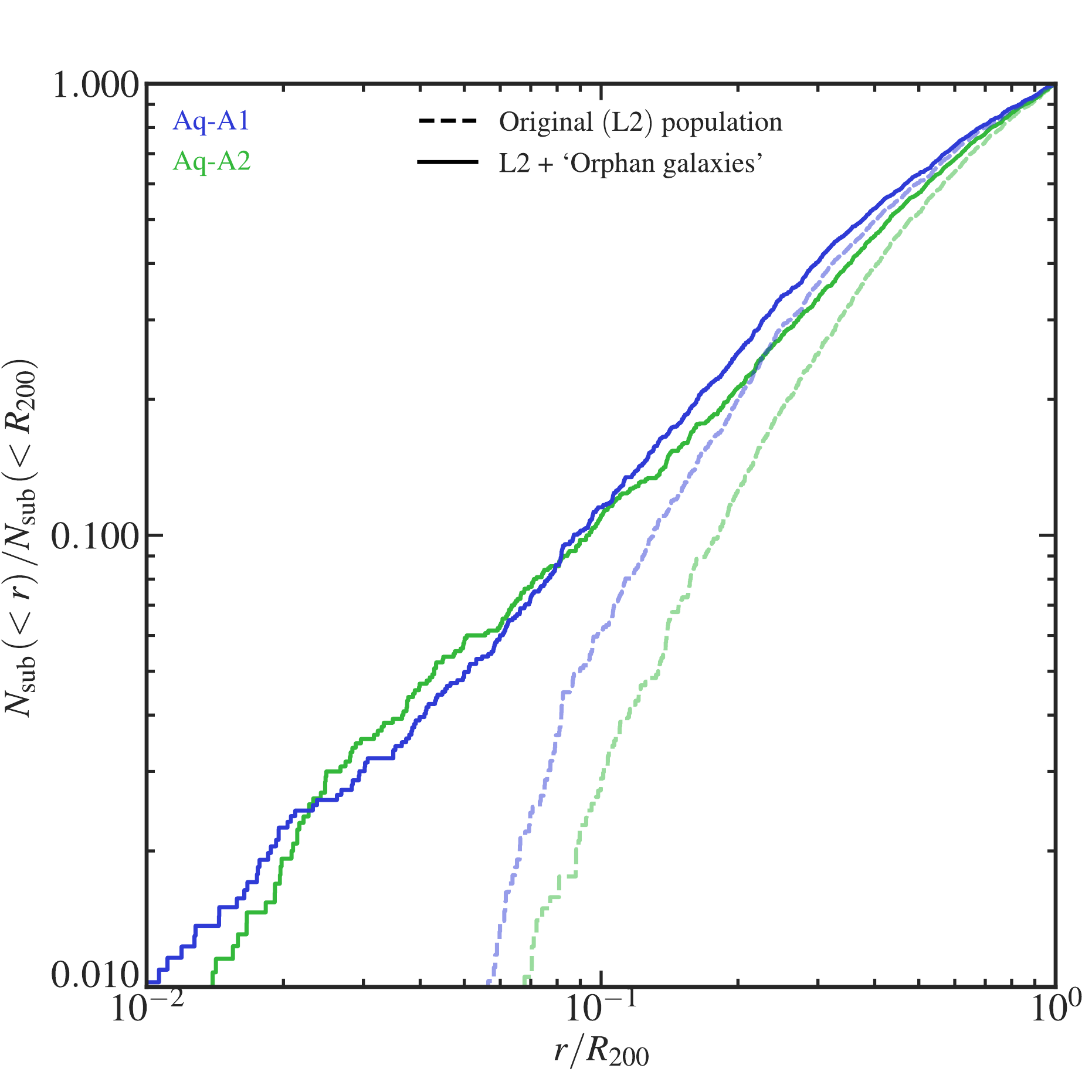}
	\vspace{-17pt}
	\caption{Normalized cumulative subhalo number counts for the Aq-A1 and Aq-A2 haloes. The dashed lines show the original, uncorrected number counts prior to the application of \Galform{}. The solid lines show the number counts for each halo after adding `orphan galaxies' to the original population. The subhalo populations before the correction are poorly sampled in the innermost regions, and are not well-converged between the two haloes.}
	\label{fig:t2_correction_comparison}
	\vspace{-10pt}
\end{figure}

In this Appendix we provide details of the scheme that we implement to supplement the $z=0$ subhalo population of each \Aquarius{} halo with subhaloes that are otherwise unresolved at this time. We also compare the difference these additions make to the subhalo number density profile.

The semi-analytic model \Galform{} described by \citet{lacey_unified_2016}, which is based on the same cosmology as the \Aquarius{} simulation suite, is applied to each of the \Aquarius{} DM haloes in turn. We use the \citet{simha_modelling_2017} merging scheme to track the dynamical evolution of subhaloes over the course of cosmic time. Well-resolved subhaloes are tracked directly by the N-body simulation; however, those that fall below the resolution limit are lost. \citeauthor{simha_modelling_2017} recover this population by tracking the most bound particle in these subhaloes from the last epoch at which they were associated with a resolved subhalo. They then remove subhaloes from this population if one of the following criteria is satisfied:
\begin{enumerate}
	\item A time has elapsed after the last epoch at which the subhalo was resolved, which is equal to or greater than the dynamical friction timescale.
	\item The subhalo passes within the halo tidal disruption radius at any time.
\end{enumerate}
In both of the above cases the effects of tidal stripping on the subhalo are ignored, as are interactions between orbiting subhaloes.

In \figref{fig:t2_correction_comparison} we compare the normalized cumulative radial subhalo counts of the \Aquarius{} A1 and A2 haloes with the $\vpeak{}\geq\kms[10]$ selection threshold applied. Prior to the application of \Galform{} the original normalized subhalo counts are highly discrepant in the inner regions of the haloes. The spread in the predicted counts at $\MV{}=0$ in Aq-A1 and Aq-A2 is also wider than the spread in predictions from the other L2~haloes~(B2--E2). When correcting for the `orphan' population, which is very centrally concentrated, the discrepancy in the Aq-A1 and Aq-A2 normalized subhalo counts is almost completely eliminated. As a result the spread in the $\MV{}=0$ predictions is also reduced such that it is much smaller than the spread in the predictions from the other `L2~+~orphans' haloes. The spread in these latter predictions is also significantly reduced by the correction, which shows that failing to account for this artificially inflates the halo-to-halo scatter.
\FloatBarrier
\section{Baryonic Effects}
\label{app:Baryonic_effects}
\FloatBarrier

\begin{figure}
	\includegraphics[width=\columnwidth]{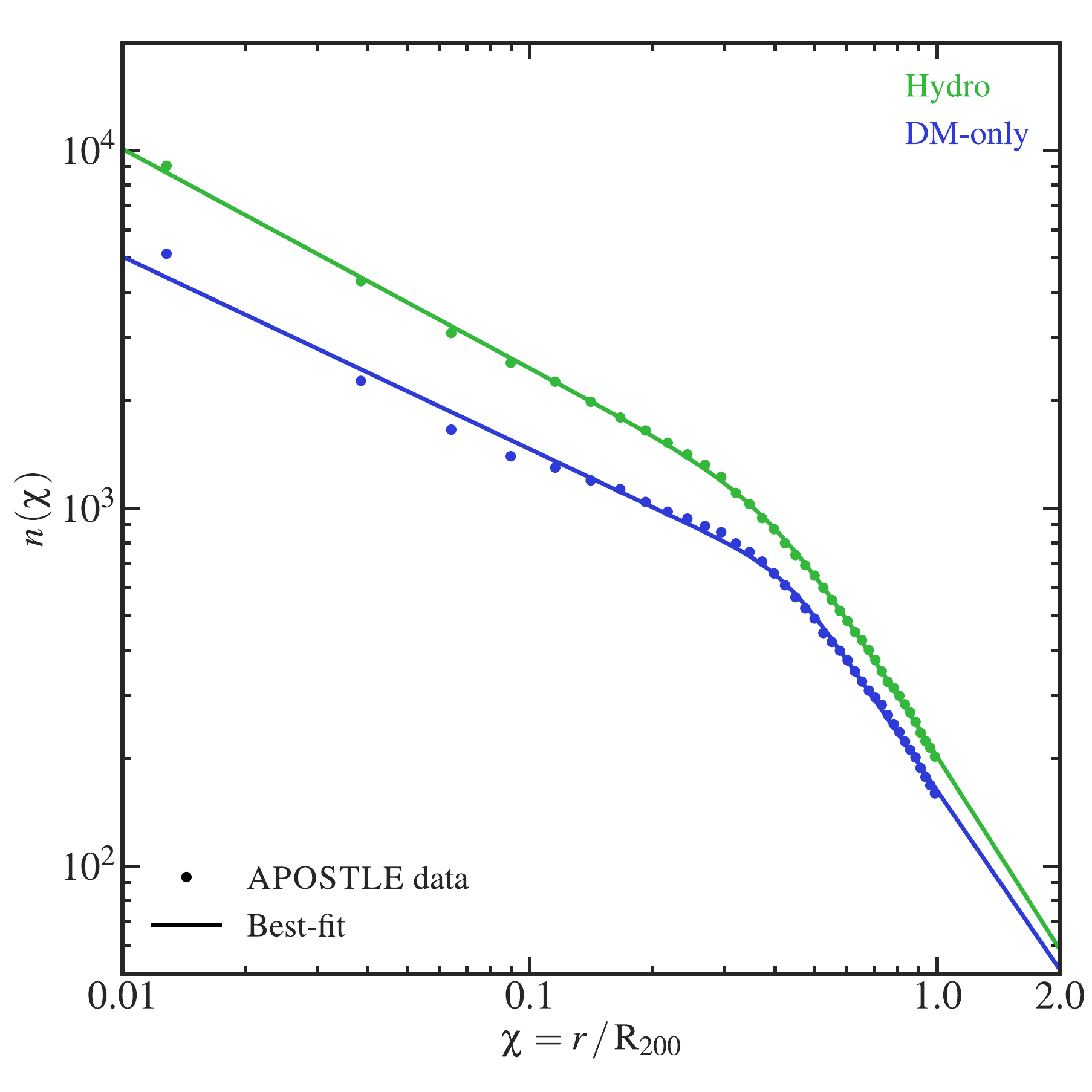}
	\vspace{-17pt}
	\caption{Fits to subhalo number density profiles in DM-only and hydrodynamic simulations. The points show averaged radial profiles for four \Apostle{} haloes. To obtain better statistics, these points were also averaged over \Gyr{5} of cosmic time; see \citet{sawala_shaken_2017} for details. The solid lines show the best-fitting double power-laws (see the main text for the best-fitting parameters).}
	\label{fig:sawala_density_fits}
	\vspace{-10pt}
\end{figure}

\citet{donghia_substructure_2010},\citet{sawala_shaken_2017}, and \citet{garrison-kimmel_not_2017} identify systematic differences in the subhalo radial number density profiles of haloes in DM-only and hydrodynamic simulations. The enhanced tidal stripping by the central baryonic disc leads to a reduction in the number of subhaloes in hydrodynamic simulations compared to their DM-only counterparts. The subhalo depletion is a radially varying function that peaks in the innermost regions of the host halo.

The subhalo number density profiles can be fit using a double power-law functional form, which is given in \citet[][equation~2]{sawala_shaken_2017}. With help from Till Sawala (private communication), we determined that some of the values stated for the fitting parameters of equation~(2) in the published version of the paper are incorrect. Taking the raw data from Till Sawala, we made our own fits, binning the data in units of $\chi=r\, /\, \Rvir{}$. \figref{fig:sawala_density_fits} gives the averaged subhalo number density profiles of four MW-like haloes from the \Apostle{} suite. To improve our statistics we also average over \Gyr{5} of cosmic time, similar to \citeauthor{sawala_shaken_2017} To these profiles, we fit a double power-law of the form
\begin{equation}
\rho\left(r\right) = 2^{\left(\beta-\gamma\right)/\alpha}\rho_s \left(\conc{}\chi\right)^{-\gamma} \left(1 + \left[\conc{}\chi\right]^\alpha\right)^{\left(\gamma-\beta\right)/\alpha},
\end{equation}
which gives fitting parameters of
$$\left(\conc{},\rho_s,\alpha,\beta,\gamma\right)=\left(2.50, 875, 4.41, 1.80, 0.613\right)$$
and
$$\left(\conc{},\rho_s,\alpha,\beta,\gamma\right)=\left(2.35, 613, 8.35, 1.66, 0.537\right)$$ for the DM-only and hydrodynamic simulations, respectively.

These fits are only constrained in the radial range $\left[0.01,1.0\right]\chi$ but in practice we extrapolate the profiles over a slightly wider range of $\left[10^{-3},2.0\right]\chi$ to subsample our haloes. We find that only minimal extrapolation is required to achieve this, and that the ratio in this extended range is also slowly varying.

The subhalo depletion is given by the ratio between the hydrodynamic and DM-only subhalo number density profiles. We compute this using the best-fitting double power-law fits given above. The ratio varies from ${\sim}0.5$ for the inner halo to about ${\sim}0.8$ at \Rvir{}. We correct the \Aquarius{} subhalo distributions using this depletion value. For each subhalo, we compute the subhalo depletion value at its radial position and use a Monte Carlo approach to decide if this subhalo is retained or discarded. Only retained subhaloes are used as input to the Bayesian inference method.
\section{Testing previous methods}
\label{app:Testing_prev_methods}

Here, we test the \citetalias{tollerud_hundreds_2008} method by applying it to a
set of mock satellite observations. This is similar to the exercise in
\secref{sec:Validation}, where, using the same blind mock observations, we
demonstrated that the Bayesian approach introduced in this paper successfully
infers the input `true' luminosity function used to generate the mock
observations.

\begin{figure}
	\includegraphics[width=\columnwidth]{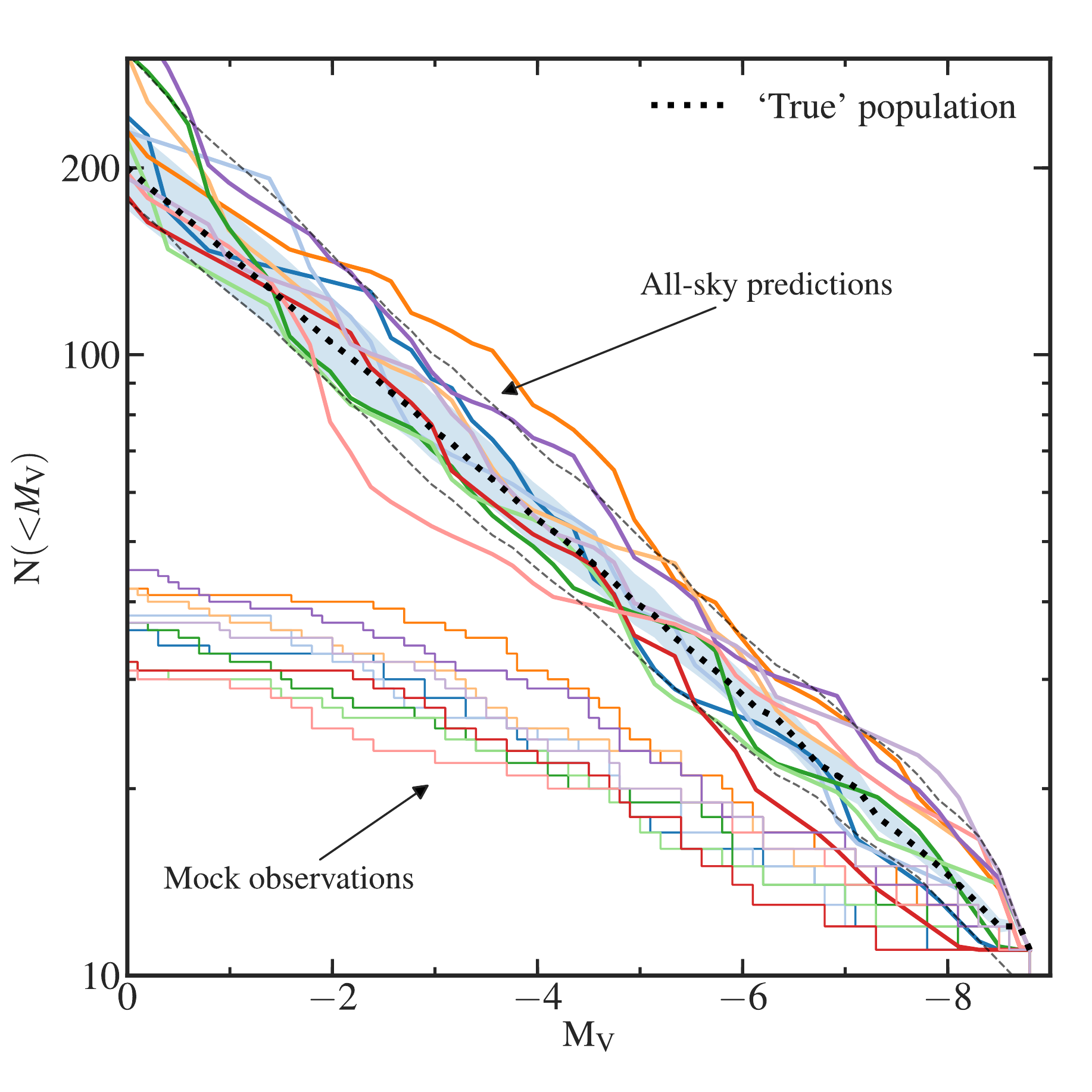}
	\vspace{-17pt}
	\caption{Test of the \citetalias{tollerud_hundreds_2008} method using mock
	observations. The thick dotted line shows the input luminosity function
	used to create the $10$ \SDSS{} mock observations, whose luminosity
	functions are shown as thin solid lines. Each of the mock observations was
	used, in turn, to predict a cumulative satellite luminosity function, with
	the corresponding results shown as thick solid lines. The shaded region
	represents the $68$~per cent~(statistical) uncertainty from one of the mocks,
	shifted to lie on top of the input luminosity function. The dashed lines
	bound the $68$~per cent~(statistical) confidence region over the medians of all
	$100$ mock predictions.}
	\label{fig:Modified_T08_validation}
	\vspace{-10pt}
\end{figure}

A set of $100$ mock \SDSS{} observations was generated from a `true' population
by one of the authors (MC; see \secref{sec:Validation} for a description of the
mocks) and supplied to another (ON), who applied the
\citetalias{tollerud_hundreds_2008} method. In order to return an unbiased
estimate, we applied the \citetalias{tollerud_hundreds_2008} approach using a
completeness radius that corresponds to a detection efficiency, $\epsilon=0.5$,
and used as input only those observed satellites with detection efficiencies,
$\epsilon \geq 0.5$. Using a random sample of $10$ mock observations, we
compare in \figref{fig:Modified_T08_validation} the scatter among the various
mocks with the typical error of the \citetalias{tollerud_hundreds_2008} method.
We find that the typical $68$~per cent~(statistical) uncertainty range estimated by the
\citetalias{tollerud_hundreds_2008} method is too low: for most magnitude
values, most of the $10$ mocks are outside the $68$~per cent~(statistical) confidence
interval. This was also demonstrated in \figref{fig:aq_extrap_methods} and
arises because the \citetalias{tollerud_hundreds_2008} method does not
incorporate the effects of stochasticity into its estimation of the
uncertainties.
%\FloatBarrier

\FloatBarrier
\pagebreak
\section{Data table}
\label{app:Data_table}
\FloatBarrier
\begin{table}
	\centering
	\caption{Cumulative number of satellites as a function of absolute magnitude
	within a heliocentric distance of \Rout{} for a \Msun[1.0\times10^{12}] MW
	halo, inferred from a Bayesian analysis of the \SDSSDR{9} + \DES{} observed
	satellites. The cumulative number of these observed satellites is provided
	for reference. The quoted confidence limits are for statistical errors
	only.}
	\label{tab:datatable}
	\begin{tabular}{@{}rccccc@{}} % four columns, alignment for each
		\hline
		\multirow{2}{*}{${\rm M_V}$} & \multicolumn{2}{c}{${\rm N}\left(<\MV{}\right)$} & \multicolumn{3}{c@{}}{Confidence limits: lower -- upper} \\
		& Observed & Predicted & $68\%$ & $95\%$ & $98\%$ \\
		\hline
		$-8.8$ & $11$ & $11$ & \ldots & \ldots & \ldots \\
		$-8.5$ & $12$ & $13$ & $12 - 15$ & $12 - 19$ & $12 - 21$ \\
		$-8.0$ & $12$ & $14$ & $13 - 16$ & $12 - 20$ & $12 - 21$ \\
		$-7.5$ & $12$ & $15$ & $13 - 17$ & $13 - 21$ & $13 - 22$ \\
		$-7.0$ & $12$ & $15$ & $14 - 17$ & $13 - 21$ & $13 - 23$ \\
		$-6.5$ & $13$ & $16$ & $14 - 19$ & $13 - 23$ & $13 - 25$ \\
		$-6.0$ & $14$ & $19$ & $16 - 22$ & $15 - 27$ & $15 - 30$ \\
		$-5.5$ & $16$ & $22$ & $19 - 26$ & $17 - 32$ & $16 - 34$ \\
		$-5.0$ & $18$ & $27$ & $23 - 32$ & $20 - 39$ & $20 - 43$ \\
		$-4.5$ & $20$ & $31$ & $27 - 38$ & $23 - 47$ & $22 - 50$ \\
		$-4.0$ & $23$ & $41$ & $35 - 49$ & $30 - 60$ & $29 - 64$ \\
		$-3.5$ & $30$ & $52$ & $44 - 62$ & $39 - 76$ & $37 - 82$ \\
		$-3.0$ & $33$ & $61$ & $51 - 73$ & $44 - 89$ & $43 - 95$ \\
		$-2.5$ & $37$ & $77$ & $64 - 93$ & $55 - 114$ & $52 - 123$ \\
		$-2.0$ & $39$ & $89$ & $74 - 108$ & $63 - 133$ & $60 - 142$ \\
		$-1.5$ & $41$ & $96$ & $79 - 118$ & $67 - 147$ & $63 - 158$ \\
		$-1.0$ & $41$ & $105$ & $86 - 131$ & $72 - 163$ & $68 - 175$ \\
		$-0.5$ & $41$ & $115$ & $92 - 146$ & $75 - 186$ & $71 - 203$ \\
		$0.0$  & $42$ & $124$ & $97 - 164$ & $78 - 225$ & $73 - 249$ \\
		\hline
	\end{tabular}
\end{table}
\FloatBarrier

%%%%%%%%%%%%%%%%%%%%%%%%%%%%%%%%%%%%%%%%%%%%%%%%%%

% Don't change these lines
\bsp	% typesetting comment
\label{lastpage}
\end{document}